\documentclass[journal,10pt]{IEEEtran} 

\usepackage{times,amsmath,amssymb,amsfonts,algorithm,epsfig,subfigure,cite}
\usepackage{color} 
\UseRawInputEncoding 
\usepackage[algo2e]{algorithm2e}
\usepackage{multirow}
\usepackage{makecell}
\usepackage{color}
\usepackage{enumitem}
\begin{document}
\title{QoE-Driven Secure Video Transmission in Cloud-Edge Collaborative Networks}


\author{Tantan Zhao, Lijun He, Xinyu Huang, Fan Li 
	\thanks{
		Copyright (c) 2021 IEEE. Personal use of this material is permitted. However, permission to use this material for any other purposes must be obtained from the IEEE by sending a request to pubs-permissions@ieee.org.
		
		Manuscript received June 07, 2021; revised August 26, 2021; accepted October 25, 2021.
		The review of this paper was coordinated by Prof. Kaigui Bian. This work was supported in part by the National Natural Science Foundation of China under Grant 62071369 and in part by the Key Research and Development Program of Shaanxi Province under Grant 2020KW-009. \textit{(Corresponding author: Fan Li.)}
		
		T. Zhao, L. He, X. Huang and F. Li are with the Ministry of Education Key Laboratory for Intelligent Networks and Network Security, School of Information and Communications Engineering, Xi'an Jiaotong University, Xi'an 710049, China (e-mail: \{zhao10111446, xinyu\_huang\}@stu.xjtu.edu.cn; \{lijunhe, lifan\}@mail.xjtu.edu.cn).
}}

\maketitle
%
\begin{abstract}
Video transmission over the backhaul link in cloud-edge collaborative networks usually suffers security risks, which is ignored in most of the existing studies. The characteristics that video service can flexibly adjust the encoding rates and provide acceptable encoding qualities, make the security requirements more possible to be satisfied but tightly coupled with video encoding by introducing more restrictions on edge caching. In this paper, by considering the interaction between video encoding and edge caching, we investigate the quality of experience (QoE)-driven cross-layer optimization of secure video transmission over the wireless backhaul link in cloud-edge collaborative networks. First, we develop a secure transmission model based on video encoding and edge caching. By employing this model as the security constraint, then we formulate a QoE-driven joint optimization problem subject to limited available caching capacity. To solve the optimization problem, we propose two algorithms: a near-optimal iterative algorithm (EC-VE) and a greedy algorithm with low computational complexity (Greedy EC-VE). Simulation results show that our proposed EC-VE can greatly improve user QoE within security constraints, and the proposed Greedy EC-VE can obtain the tradeoff between QoE and computational complexity.
\end{abstract}


\IEEEpeerreviewmaketitle

\section{Introduction}
With the rapid development of video processing and mobile communication technology, large numbers of ultra-high-definition (UHD) video-on-demand (VoD) facing massive links have emerged \cite{tian2017tmc}. The number of video requests is growing dramatically, which leads to the heavy traffic load of networks and high latency of users. These factors could bring poor quality of experience (QoE) for users. Luckily, edge caching (EC) technology, as one of the key technologies of the cloud-edge collaboration networks in the fifth generation (5G) system, provides a promising method to handle these challenges \cite{hu2015etsi, abbas2018iot}. Since EC servers are closer to the users at the network edge and own powerful intelligent storage capabilities to precache hot videos, massive video contents can be delivered directly to users from EC servers to reduce backhaul traffic burden and transmission latency \cite{sun2020tvt,qu2020tmc,chen2019icme}. In addition to transmission latency, video encoding quality is also a key factor affecting QoE. Video encoding quality is influenced by video encoding parameters: video encoding rate. Higher video coding rate means lager total size of the video packets and will bring higher received video quality. Transmission latency depends on both video encoding parameters and caching strategy. This is because most of the transmission time is spent on downloading the uncached video packets over the backhaul link from MBS. Generally speaking, high video encoding quality and low transmission latency imply high QoE. While in the practical scenarios with fixed available caching capacity, too high video encoding quality may result in too high transmission latency and vice versa. To maximize QoE, we need to find the tradeoff between video encoding quality and transmission latency by jointly optimizing video encoding parameters and caching strategy subject to limited available caching capacity. Therefore, how to balance video encoding characteristics and edge caching to improve user QoE is worth studying.

Recently, the copyright protection of video is becoming more and more important. Moreover, the emerging paid VIPs and confidential video services make the secure transmission of video contents become an issue that cannot be ignored in cloud-edge collaborative networks. To be specific, for the video transmission in cloud-edge collaborative networks, when video data is transmitted from the cloud server to the user in unicast mode, the eavesdropping user and the legitimate user receive information on the same frequency channel at the same time. Once channel condition of the eavesdropper is better than that of the legitimate user, the eavesdropper could successfully intercept the video packets, which causes data leakage and infringes on video copyright and benefits of users. Therefore, secure video transmission is important. 

For the transmission security of wireless links from the EC servers to the users, the existing physical layer secure (PLS) mechanism, which has been studied intensively and extensively \cite{dong2010tsp,yang2015cm,liu2017cst,chen2018tvt}, is sufficient.  However, the PLS mechanism is complicated because it could ensure the security of each video packet. If it is utilized to ensure the transmission security of wireless backhaul links in the distributed cache scenarios, where an EC server may be connected to multiple base stations, the system overhead could be unbearable. Therefore, it is an urgent issue to develop a secure mechanism with low complexity to ensure secure transmission of wireless backhaul links, rather than using the strict and complicated traditional PLS mechanism.

The flexible and cost-effective wireless backhaul has become an attractive alternative to wired backhaul for small cell networks. However, wireless backhaul is vulnerable to wiretapping due to the intrinsic broadcasting nature of wireless channels. Therefore, the issue of secure backhaul links has aroused wide interest of researchers in recent years. Considering unreliable wireless backhaul connections, the authors in \cite{zhang2019tvt} investigated the secrecy performance of cooperative single carrier HetNets. The authors in \cite{yen2017access} studied the effect of unreliable wireless backhaul links on the secrecy performance of an energy harvesting relay network. Different transmitter selection schemes were proposed to enhance the secrecy. Wireless backhaul along with secrecy was considered for CR network in \cite{yin2018access} and \cite{ong2018iet}. 

However, most of the existing studies related to secure backhaul links only analyzed the significant negative impact of the backhaul unreliability on secure transmission instead of developing methods to enhance the reliability of wireless backhaul links. Furthermore, the issue of the secure backhaul links in the cloud-edge collaborative networks is rarely considered. The authors in \cite{Gabry2016icc} considered the security risks of the backhaul links when studying the optimal cache placement in edge heterogeneous networks. Nevertheless, the issue aimed at general data transmission, ignoring the characteristics of the video applications, which resulted in the poor performance for video transmission applications. Therefore, secure video transmission through the backhaul links in the cloud-edge collaborative networks is still an open question.

To address the abovementioned problems in secure video transmission via backhaul links in cloud-edge collaborative networks, we propose a QoE-driven cross-layer optimization algorithm to improve video encoding quality, reduce transmission latency and guarantee security of backhaul links. The main contributions of this paper can be summarized as follows. 

\begin{enumerate}[fullwidth,itemindent=1em]
	

\item \textit{Problem Formulation of Joint Optimization of Video Encoding and Edge Caching:} Guaranteeing secure video transmission is of equal importance in the cloud-edge collaborative network besides optimizing QoE. Since video service can flexibly provide different acceptable encoding qualities to adapt to the real scenario's requirements by adjusting the encoding parameters correspondingly, the extra security requirements can be satisfied with higher possibility but has to be tightly coupled with video encoding by introducing more restrictions on edge caching. Therefore, how to find the tradeoff among transmission latency, video encoding quality and security becomes the new challenge. To address this issue, we first develop a secure transmission model based on video encoding and edge caching. By employing this model as the security constraint, we comprehensively consider the interaction of video encoding and edge caching and then formulate a QoE-driven joint optimization problem. The optimization problem aims to optimize both video encoding and caching strategies simultaneously subject to the limited available caching capacity. 

\item \textit{Near-Optimal and Suboptimal Solutions:} The formulated joint optimization problem is a high-dimension non-linear mixed integer programming problem, which makes it more difficult to achieve analytical solution. On one hand, the non-smooth objective function, the non-convex constraints and multiplication among the optimization variables make the original optimization problem intractable. On the other hand, the high dimensionality of optimization variables makes it difficult to obtain the optimal solution. Therefore, we propose a near-optimal algorithm to solve it. To be specific, efficacious transformations are performed on the original intractable problem to obtain its relaxing solution. Then, Boolean bound is developed to obtain the near-optimal solution by iteration until all Boolean variables are integers. Furthermore, considering that the near-optimal algorithm is time-consuming, we also propose a suboptimal algorithm based on the greedy method with low computational complexity to get the tradeoff between the user QoE and the computational complexity.   

\end{enumerate}

The rest of this paper is organized as follows. Section \uppercase\expandafter{\romannumeral2} reviews the related works. In Section \uppercase\expandafter{\romannumeral3}, we introduce the system model, which includes the framework of QoE-driven cross-layer system and describes the secure transmission model. In Section \uppercase\expandafter{\romannumeral4}, we formulate the video encoding quality maximization and latency minimization problem under the secure constraints. In Section \uppercase\expandafter{\romannumeral5} we propose a near-optimal solution based on joint optimization of video encoding and edge caching. In Section \uppercase\expandafter{\romannumeral6}, we propose a suboptimal solution based on greedy method. Section \uppercase\expandafter{\romannumeral7} gives the experimental results, and discusses the performance gains of the proposed algorithms compared to the existing algorithms. Finally, Section \uppercase\expandafter{\romannumeral8} concludes this paper. 

\section{Related Work}

In this section, we will review and analyze the progress of existing related work from the following two aspects.
\subsection{QoE-Driven Edge Video Caching}

The edge servers are closer to users and have powerful storage capabilities. By caching multimedia contents during off-peak hours, delay and congestion can be reduced \cite{ren2019tvt}, thereby improving mobile user QoE \cite{mao2017cst,shi2016iot}. 

To exploit the storage capacity of edge servers, the authors in \cite{li2015tc} proposed a distributed cache optimization algorithm through belief propagation of heterogeneous networks, which greatly reduced the download delay of users. From the perspective of the basic information theoretical limit of EC, the authors in \cite{Sengupta2016ciss} gave the best trade-off between latency and caching. Considering the particularity of video transmission, the researchers in \cite{Hsu2016cl} proposed a co-caching strategy for small-cell base stations (SBSs) and mobile devices, which greatly reduced the delay in video content delivery. 

In addition to latency, the popularity of video contents also determines whether the cached videos meet the needs of users. The authors in \cite{Müller2017twc, li2016tm} proposed a context-aware caching scheme to predict the popularity and learn the video popularity of a specific context online, which is used to determine cache replacement strategy. However, with the increasing heterogeneity of user groups demanding specific video content, the issue of caching mobile video streaming has become a more complicated task. Therefore, there are also many papers that use machine learning to predict preferences of users \cite{hou2018ics, wang2015tpds, zhang2019wc, Saputra2019wcl, ale2019iot}. 

The studies mentioned above are for traditional video streaming, to improve the QoE of dynamic adaptive streaming users. The authors in \cite{Li2018tm} studied the QoE-driven optimization of EC placement for dynamic adaptive video streaming, which considers the different rate-distortion characteristics of videos and the coordination among distributed EC servers. The authors in \cite{huang2020icme} proposed a refined caching update strategy based on video popularity, content importance, and user playback status, which ensured that the video segments to be played by users could be cached in time. To further improve user QoE, multiple modules in addition to EC are considered to be optimized comprehensively. The authors in \cite{Liang2017twc} proposed to enhance the QoE-aware wireless edge caching with bandwidth provisioning in software-defined wireless networks. Specifically, a joint optimization mechanism of caching strategy, bandwidth configuration, and adaptive video streaming was designed to reduce delay and improve QoE. In addition, the researchers in \cite{huang2020twc} proposed a joint scheme of caching strategy, power allocation, user association, and adaptive video streaming to improve the system spectrum efficiency and user QoE. However, the scheme in \cite{huang2020twc} will cause a greater burden on backhaul traffic. Therefore, in \cite{Mehrabi2019tgcn}, the authors studied the joint optimization scheme of mobile terminal QoE and backhaul traffic, which reduced power consumption and backhaul traffic, meanwhile improving QoE. Although the above studies aim to improve the QoE of mobile users, security issues in the cloud-edge collaborative networks were not considered.
\vspace{-0.2cm}
\subsection{Caching-Based Secure Transmission}

With the increasing complexity of network environment, legal transmission in the networks is inevitably attacked by external malicious nodes, or the SBS originally used for caching becomes a malicious node and implements eavesdropping. Therefore, caching-based secure transmission scheme has aroused wide interest of researchers. The secure schemes proposed by these studies can be classified as two categories according to different methods used to achieve security, including encryption and coding based secure schemes and wireless security mechanism based secure schemes.


For the encryption and coding based secure scheme, in \cite{Zewail2020tifs}, the authors studied the secure device-to-device coding and caching scheme in the sense of information theory. The authors in \cite{Kiskani2017tvt} proposed a decentralized secure coding caching method in wireless ad-hoc networks. The researchers in \cite{Xiao2018wc} studied the attack models in EC systems and proposed lightweight authentication and secure collaborative caching schemes based on reinforcement learning techniques for the mobile edge caching. The authors in \cite{su2018tii} proposed a secure caching solution for disaster backup in mobile social networks using fog computing. In \cite{xu2019itj}, a secure caching scheme in heterogeneous networks for multi-homing users is designed. A trust mechanism and a Chinese remainder theorem-based privacy preservation protocol is proposed.

For the wireless security mechanism based secure scheme, in \cite{xiangtwc2018}, a caching scheme to enhance PLS for a backhaul-limited cellular network was proposed. The authors focused on minimizing the transmit power subject to the secrecy rate, where the backhaul links are considered as wired. In \cite{zhengtc2018}, the authors studied the secure transmission problem in a cache-assisted heterogeneous networks. To realize the secure and energy-efficient transmission, a joint cache placement and file delivery scheme was proposed. In \cite{haoiot2020}, the authors investigated the secure transmission delay minimization problem in an edge cache-assisted millimeter-wave cloud radio access network. The authors designed beamforming to minimize the fronthaul transmission delay and access transmission delay.

Apart from the above studies, for the security risks of backhaul links, the authors in \cite{Gabry2016icc} studied the optimal cache placement problem under secrecy constraints of backhaul links in edge heterogeneous networks. Specifically, the authors proposed an edge caching strategy to minimize the average backhaul links rate and ensure the security of backhaul links. However, what has been studied in this study is a secure transmission scheme for general data. The unique characteristics of videos, which allow distortion within acceptable range, is ignored. Generally speaking, different encoding parameters correspond to different encoding qualities, which bring different QoE levels. Therefore, the proposed secure scheme is not specific to video transmission.

\section{SYSTEM MODEL}
\subsection{Framework}


\begin{figure}[!t]
	\centering
   	\includegraphics[width=80mm]{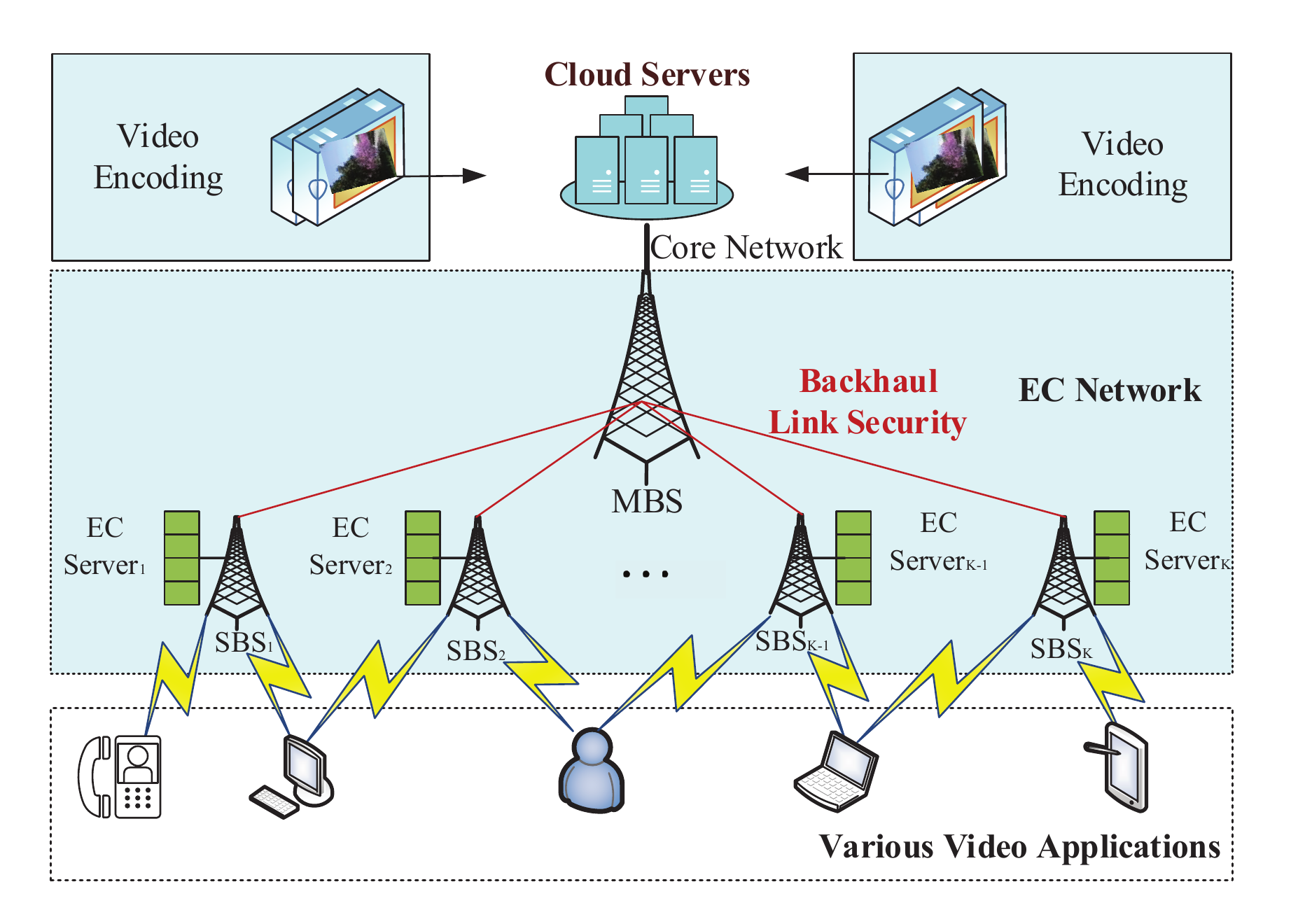}
	\caption{VoD downlink transmission in cloud-edge collaborative networks.}
	\label{fig_system_model}
\end{figure}

\begin{figure}[!t]
	\centering
	\includegraphics[width=80mm]{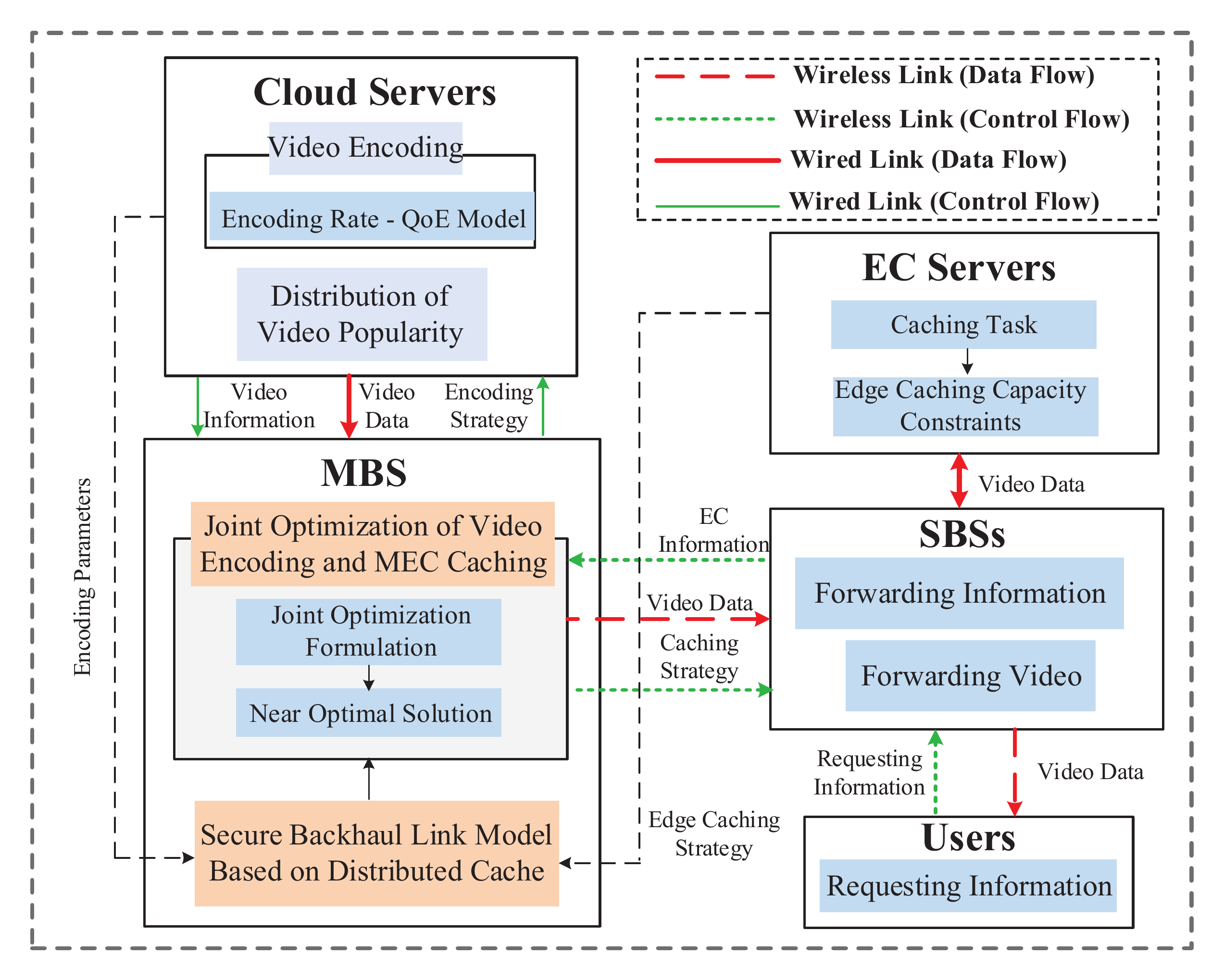}
	\caption{QoE-driven joint optimization scheme of video encoding and edge caching based on secure backhaul links model.}
	\label{fig_scheme_model}
\end{figure}

In this paper, we consider a VoD downlink scenario in the cloud-edge collaborative networks, where multiple users request video files from the cloud servers via a macro-cell base station (MBS), which has access to the cloud servers through wired core network. SBSs are deployed in the coverage area of the MBS to serve user video requests and communicate with the MBS via wireless backhaul links. As shown in Fig. \ref{fig_system_model}, the solid red lines denote the wireless backhaul links between the MBS and SBSs, following\cite{sahajsac2019, h2020tvt, yangiot2021}. The number of EC servers is $K$, and let ${\cal K}$ denote the set of $K$ EC servers and the caching capacity of each EC server is ${\Phi _k}, \forall k \in {\cal{K}}$. Each EC server is connected to an SBS. A user requesting video files is initially served by SBSs. For video services without packet loss, $n$ encoding packets are required to successfully decode video information. If the number of video packets sent by all EC servers within the connection range is $m$ and less than $n$, the decoding conditions are not met. Then, the MBS is contacted to send the remaining $n-m$ video packets to the user via backhaul links. Based on the above scenario, we propose a QoE-driven joint optimization scheme of video encoding and edge caching based on the secure transmission model for the backhaul links, as illustrated in Fig. \ref{fig_scheme_model}. 

Specifically, the whole block diagram of the joint optimization scheme contains the modules of MBS, cloud servers, SBSs, EC servers and users. First, the video encoding information, i.e., encoding rate, QoE model and the distribution of popularity, from the cloud servers, and the EC information, i.e., the caching capacity and the number of EC servers from the SBSs, are sent to the system analysis module MBS. Then, the optimal encoding and caching strategy based on the collected information and the secure transmission model of backhaul links are designed by the MBS and sent to the cloud servers and the SBSs, respectively. The secure transmission model of the backhaul links is based on video encoding and edge caching. This means that the encoding packet size, the number of encoding packets cached on the EC servers and the capacity of EC servers together determine whether the secure transmission can be achieved. Finally, the video streamings are encoded by the cloud servers according to the designed encoding strategy and cached on the EC servers according to the designed caching strategy after being transmitted from the cloud servers. Here we consider utilizing the adjustable encoding parameters and the distributed edge cache to avoid data leakage over the wireless backhaul links from the MBS to the SBSs and construct the secure transmission model of wireless backhaul links. Based on the secure transmission model, the QoE-driven joint optimization model of video encoding and edge caching is established. 
\vspace{-0.2cm}
\subsection{Secure Transmission Model Based on Video Encoding and Edge Caching}
\setlength{\belowdisplayskip}{3pt}
For the high-definition (HD)/UHD VoD services, we assume that there are $F$ original video streamings to be encoded and cached, and let $\cal{F}$ denote the set of $F$ video streamings. ${p_j}, \forall j \in {\cal{F}}$ indicates the requested probability of the $j$-th video file, which is ranked in descending order according to video popularity. Without loss of generality, we can mark the most popular video file with index $1$ and the least popular video file with index $F$. Specifically, the requested probability of a video file can be described as a Zipf distribution. Assuming that all video files are sorted according to their popularity, the probability of being requested for each video file can be obtained by the following formula:
\begin{equation}
	{p_j} = \frac{{{1 \mathord{\left/
					{\vphantom {1 {{j^{ - \theta }}}}} \right.
					\kern-\nulldelimiterspace} {{j^{ - \theta }}}}}}{{\sum\limits_{j = 1}^F {{1 \mathord{\left/
						{\vphantom {1 {{j^{ - \theta }}}}} \right.
						\kern-\nulldelimiterspace} {{j^{ - \theta }}}}} }},0 < \theta  < 1, \forall j \in {\cal{F}}
\end{equation}
where $\theta$ is the tilt factor of the distribution, which is used to control the distribution of video popularity, and the larger value means that the requested probability is more concentrated.  

To be specific, the following assumptions are made when encoding and caching video streamings: 

\begin{enumerate}[fullwidth,itemindent=1em]

\item Each video file is encoded into $n$ packets of different sizes, and let $\cal{N}$ denote the set of $n$ encoding packets. Let ${s_{i,j}}$ denote the size of the $i$-th video packet of the $j$-th video file. The $n$ video encoding packets are cached distributedly on different EC servers, following \cite{Liang2017twc} and \cite{zhao2017tmm}.

\item ${\tilde m_{k,i,j}}$ denotes whether the $i$-th video packet of the $j$-th video file is 
cached on the $k$-th EC server $\left( {{{\tilde m}_{k,i,j}}{\rm{ = 1}}} \right)$ or not $\left( {{{\tilde m}_{k,i,j}}{\rm{ = 0}}} \right)$.

\end{enumerate}

Based on the above assumptions, the cooperative caching strategy among multiple EC servers can be formulated as a matrix ${{\bf{M}}_{K \times F}}$ of size $K \times F$, where the rows indicate the EC servers, the columns indicate the video files, and the element ${m_{k,j}}$ represents the number of encoding packets of the $j$-th video file cached on the $k$-th EC server, obviously, ${m_{k,j}} = \sum\limits_{i = 1}^n {{{\tilde m}_{k,i,j}}}$. As a result, for each request, the number of encoding packets of the $j$-th video file that need to be transmitted through the backhaul links can be expressed as 
\begin{equation}
	R_j = n - \min \left( {n,\sum\limits_{k = 1}^K {{m_{kj}}} } \right)
	\label{equ3}
\end{equation}

Meanwhile, assuming that the number of total requests during the whole delivery phase received by SBSs is $\Psi $, then the number of requests for the video file $j$ received by SBSs is ${\Psi _j} = \Psi  \cdot {p_j}, \forall j \in {\cal{F}}$. Therefore, for ${\Psi _j}$ requests, the total number of encoding packets for the $j$-th video file transmitted through the backhaul links can be obtained as follows:
\begin{equation}\small
	P_j = {\Psi _j}\cdot \left[ {n - \min \left( {n,\sum\limits_{k = 1}^K {{m_{kj}}} } \right)} \right]
	\label{equ6}
\end{equation}

For clarity, considering that $n \ge \sum\limits_{k = 1}^K {{m_{kj}}}$ in our considered scenario, Eq. \eqref{equ6} can be further formulated as
\begin{equation}\small
	P_j = {\Psi _j}\cdot \left( {n - \sum\limits_{k = 1}^K {{m_{kj}}} } \right)
	\label{equ7}
\end{equation}

We know that for the video services without packet loss, secure transmission can be realized if and only if $P_j \le n - 1$. Consequently, the secure transmission model of backhaul links based on the distributed edge cache can be expressed by 
\begin{equation}
	{\Psi _j}\cdot\left( {n - \sum\limits_{k = 1}^K {{m_{kj}}} } \right) \le n - 1
\end{equation}

After further simplification, we can get
\begin{equation}
	\sum\limits_{k = 1}^K {{m_{kj}}}  \ge n\cdot\left( {1 - \frac{1}{{{\Psi _j}}}} \right) + \frac{1}{{{\Psi _j}}}
\end{equation}

Based on the above derivation, we know that the probability of secure transmission for the requested video is closely related to the encoding packets cached on EC servers and the number of video requests.

\section{Problem Formulation}
Based on the secure transmission model achieved for backhaul links, the QoE-driven joint optimization problem of video encoding parameters and edge caching strategy under the constraints of edge caching capacity and video encoding rate has been formulated.  We define the optimization variables as ${\bf{\tilde {\bf{M}}}}_{K \times n \times F} = \left\{{{\tilde m}_{k,i,j}}, \forall k \in {\cal{K}}, \forall i \in {\cal{N}}, \forall j \in {\cal{F}} \right\}$ and ${\bf{S}}_{n \times F} = \left\{{{s}_{i,j}}, \forall i \in {\cal{N}}, \forall j \in {\cal{F}} \right\}$, where ${{{\tilde m}_{k,i,j}}}$ and ${{s_{i,j}}}$ denote the caching strategy and the size of each video encoding packet, respectively. To improve user QoE, we construct the joint optimization problem expressed by Eq. \eqref{equ10} as follows:
\begin{subequations}
	\begin{align}
		\mathop {\max }\limits_{{\bf{\tilde {\bf{M}}}}, {\bf{S}}}& Q\left( {\bf{\tilde {\bf{M}}}}, {\bf{S}} \right)\\
		\text{s}\text{.t}\text{.}&\sum\limits_{j = 1}^F {\sum\limits_{i = 1}^n {{{\tilde m}_{k,i,j}}{s_{i,j}}} }  \le {\Phi _k},\forall k \in {\cal{K}}\label{7b}\\
		&\sum\limits_{k = 1}^K {\sum\limits_{i = 1}^n {{{\tilde m}_{k,i,j}}} }  \le n,\forall j \in {\cal{F}}\label{7c}\\
		&\sum\limits_{k = 1}^K {\sum\limits_{i = 1}^n {{{\tilde m}_{k,i,j}}} }  \ge n\cdot\left( {1 - \frac{1}{{{\Psi _j}}}} \right) + \frac{1}{{{\Psi _j}}},\forall j \in {\cal{F}}\label{7d}\\
		&{R_{{{\min }_j}}} \le \frac{{\sum\limits_{i = 1}^n {{s_{i,j}}} }}{{{T_d}}} \le {R_{{{\max }_j}}},\forall j \in {\cal{F}}\label{7e}\\
	    &{{\tilde m}_{k,i,j}} \in \left\{ {0,1} \right\},\forall k \in {\cal{K}},\forall i \in {\cal{N}},\forall j \in {\cal{F}}\label{7f}
	\end{align}
	\label{equ10}
\end{subequations}
where $Q\left(  \cdot  \right)$ is the objective function for different application demands, which can represent a single optimization objective or the combination of multiple optimization objectives, depending on the practical application demands. ${R_{{{\min }_j}}}, \forall j \in {\cal{F}}$ and ${R_{{{\max }_j}}}, \forall j \in {\cal{F}}$ denote the maximum and minimum encoding rates of the $j$-th video file, and the larger encoding rate means the higher QoE. $T_d$ denotes the duration of each video file.

To be specific, in our considered VoD downlink scenario, the QoE is determined by video encoding quality and transmission latency. The optimization objective is to maximize the video encoding quality and minimize the transmission latency simultaneously. Then, the $Q\left(  \cdot  \right)$ can be defined as
\begin{equation}
	Q\left( {\bf{\tilde {\bf{M}}}}, {\bf{S}} \right) = \sum\limits_{j = 1}^F\left[{f\left( {{s_{i,j}}} \right)} {\rm{ - }}g\left( {{{\tilde m}_{k,i,j}},{s_{i,j}}} \right)\right], 
	\label{equ11}
\end{equation}
where ${f\left( {{s_{i,j}}} \right)}$ denotes the relationship between video encoding parameters and video encoding quality, which is determined by the chosen mean opinion score (MOS) model, and $g\left( {{{\tilde m}_{k,i,j}},{s_{i,j}}} \right)$ denotes the relationship among caching strategy, video encoding parameters and transmission latency. In particular, according to the MOS model in \cite{zhao2015tcsvt}, for the $j$-th video file, ${f\left( {{s_{i,j}}} \right)}$ and $g\left( {{{\tilde m}_{k,i,j}},{s_{i,j}}} \right)$ can be expressed as  
\begin{equation}
	\begin{array}{l}
		{f\left( {{s_{i,j}}} \right)}{\rm{ = }}C_{1_j}\!\!\cdot\!\!{\left(\!\! {\frac{{\sum\limits_{i = 1}^n\!\! {{s_{i,j}}} }}{{{T_d}}}}\!\! \right)^3}\!\!\!\!\! +\! C_{2_j}\!\!\cdot\!\!{\left(\!\! {\frac{{\sum\limits_{i = 1}^n\!\! {{s_{i,j}}} }}{{{T_d}}}}\!\! \right)^2}\!\!\!\!\! +\! C_{3_j}\!\!\cdot\!\!\left(\!\! {\frac{{\sum\limits_{i = 1}^n\!\! {{s_{i,j}}} }}{{{T_d}}}} \!\!\right)\!\! +\! C_{4_j}\\
	\end{array}
\end{equation}
where $C_{1_j},C_{2_j},C_{3_j},C_{4_j}, \forall j \in {{\cal F}}$ are video encoding quality parameters which can be determined by the encoding rate range of the $j$-th video file for a given MOS model.
\begin{equation}
	\begin{array}{l}
		g\left( {{{\tilde m}_{k,i,j}},{s_{i,j}}} \right) = {V_{_j}}\!\cdot\! {\left[ {\frac{{{p_j}\cdot\left( {\sum\limits_{i = 1}^n {{s_{i,j}}}  - \sum\limits_{i = 1}^n {\sum\limits_{k = 1}^K {{{\tilde m}_{k,i,j}}{s_{i,j}}} } } \right)}}{{{R_{bk}}}}} \right]^{\frac{2}{3}}}\\
	\end{array}
\end{equation}
where ${V_{{j}}}, \forall j \in {{\cal F}}$ is the weighting coefficient of transmission latency for a given MOS model. ${{R_{bk}}}$ is the transmission rate of the backhaul links. 

Constraint \eqref{7b} indicates that the total size of video packets cached in each EC server should not be greater than the edge caching capacity ${\Phi _k}, \forall k \in {\cal{K}}$. Constraint \eqref{7c} indicates that the number of video encoding packets cached in all EC servers should not be larger than the total number of video encoding packets $n$. Constraint \eqref{7d} reveals the number of video encoding packets that need to be cached in EC servers to ensure the secure transmission of backhaul links. Constraint \eqref{7e} gives the maximum and the minimum limits of encoding rate for each video streaming. Constraint \eqref{7f} denotes caching strategy as the binary variables, which means whether the $k$-th EC server caches the $i$-th packet of the $j$-th video file or not. 

By observing the formulated optimization problem in Eq. \eqref{equ10}, we can know that caching strategy and video encoding parameters cooperatively affect the security of the transmission. Specifically, the security constraint in Eq. \eqref{7d} acts on caching strategy, meanwhile Eq. \eqref{7b} reveals that the video encoding parameters interact with caching strategy. Thus, security constraint in Eq. \eqref{7d} actually acts on caching strategy and video encoding parameters simultaneously. When the capacity of EC server is limited and security constraint is strict, video encoding parameters and caching strategy can be jointly optimized and cooperatively adjusted to ensure security. More video encoding packets could be cached on EC server in proper encoding rate to satisfy the secure constraint and the limited edge caching capacity.

\section{Near Opimal Solution Based on Joint Optimization of Video Encoding and Edge Caching}

The optimization problem formulated in section \uppercase\expandafter{\romannumeral4} is an intractable problem. Because it is a high-dimension non-linear mixed 0-1 integer programming problem. First, the non-smooth objective function and the non-convex constraints make the original optimization problem intractable. Furthermore, the multiplication and coupling among the high-dimension optimization variables make the original optimization problem even more intractable. Finally, the high dimensionality of optimization variables make it difficult to obtain the optimal solution of the original optimization problem. To solve it, a near-optimal solution based on relaxation and Boolean bound is proposed in this subsection. Specifically, the whole solving procedure can be divided into two steps, i.e., relaxation of the original high-dimension non-linear mixed 0-1 integer programming problem and the near-optimal solution based on the derived Boolean bound. In the following section of this paper, we will describe them in detail.
\setlength{\abovedisplayskip}{4pt}
\vspace{-0.1cm}
\subsection{Relaxation of the Original High-Dimension Non-linear Mixed 0-1 Integer Problem}
We first relax the original mixed 0-1 integer programming problem with integer constraints into the relaxing problem without integer constraints. Then, the relaxing solution of the relaxing problem can be obtained by utilizing GlobalSearch. We can easily get the relaxing problem, which can be expressed by the following:
\begin{subequations}
	\begin{align}
		\mathop {\max\! }\limits_{{\bf{\tilde {\bf{M}}}}, {\bf{S}}} &\sum\limits_{j = 1}^F \left[{f\left( {{s_{i,j}}} \right)} {\rm{ - }}g\left( {{{\tilde m}_{k,i,j}},{s_{i,j}}} \right)\right]\\
		\text{s}\text{.t}\text{.}&\;\eqref{7b},\eqref{7c},\eqref{7d},\eqref{7e}\\
		& 0 \le {\tilde m_{k,i,j}} \le 1,\forall k \in {{\cal K}},\forall i \in {{\cal N}},\forall j \in {{\cal F}}\label{11c}
	\end{align}
	\label{equ12}
\end{subequations}
\setlength{\abovedisplayskip}{6pt}                       
Particularly, the relaxing processing is based on the fact that the optimal solution of the mixed integer programming, denoted as ${\bf{\tilde {\bf{M}}}}_{opt}$ and ${\bf{S}}_{opt}$, will not be superior to the optimal solution of its relaxing problem, which is the upper bound of the original mixed integer programming. Here, the definition of the relaxing problem can be expressed as: the objective function and remaining constraints of the original mixed integer programming problem without considering the integer constraint \eqref{7f}, which is replaced by continuous range constraint \eqref{11c}. 

The relaxing problem \eqref{equ12} above is a maximization problem, which is obtained by replacing the integer constraint \eqref{7f} in the original mixed 0-1 integer programming problem \eqref{equ10} with the continuous range constraint \eqref{11c} in the relaxing problem \eqref{equ12}. For this maximization problem, as a typical reformulation method for solving optimization problem, we consider a minimization problem equivalent to the maximization problem \eqref{equ12} as follows:
\begin{subequations}
	\begin{align}
		\mathop {\min }\limits_{{\bf{\tilde {\bf{M}}}}, {\bf{S}}} &\sum\limits_{j = 1}^F\left[ g\left( {{{\tilde m}_{k,i,j}},{s_{i,j}}} \right){\rm{ - }}{f\left( {{s_{i,j}}} \right)}\right]\\
		\text{s}\text{.t}\text{.}&\;\eqref{7b},\eqref{7c},\eqref{7d},\eqref{7e},\eqref{11c}
	\end{align}
	\label{equ14}
\end{subequations}
\setlength{\abovedisplayskip}{6pt}
The minimization problem \eqref{equ14} is actually a non-linear optimization problem. Therefore, we consider utilizing the global optimal algorithm GlobalSearch provided by Global Optimization Toolbox of MATLAB to obtain its solution. Actually, the near-optimal solution of optimization problem \eqref{equ14} can be obtained by the use of GlobalSearch algorithm, which is a kind of heuristic search algorithm. We can denote the near-optimal solution of relaxing problem \eqref{equ14} as ${\bf{\tilde {\bf{M}}}}_r$ and ${\bf{S}}_r$, and the corresponding objective function value is $Q_r$.

\subsection{Near-Optimal Solution Employing Boolean Bound}

To obtain the near-optimal solution of the original mixed integer programming, we employ the branch and bound method \cite{1998norkinmp}. 
Specifically, we can discuss the solution of the relaxing problem \eqref{equ14} in two cases. 

In case one, the solution in ${\bf{\tilde {\bf{M}}}}_r$ concerning integer-variables ${\bf{\tilde {\bf{M}}}}_{K \times n \times F} = \left\{{{\tilde m}_{k,i,j}}, \forall k \in {\cal{K}}, \forall i \in {\cal{N}}, \forall j \in {\cal{F}} \right\}$ are integers. Then, the near-optimal solution ${\bf{\tilde {\bf{M}}}}_r$ and ${\bf{S}}_r$ of the relaxing problem is also the near-optimal solution of the original non-linear mixed optimization problem and the near-optimal solution is obtained as ${\bf{\tilde {\bf{M}}}}_{opt}={\bf{\tilde {\bf{M}}}}_{r}$, ${\bf{S}}_{opt}={\bf{S}}_r$.

In case two, an element of solution ${\bf{\tilde {\bf{M}}}}_r$ concerning integer-variables ${\bf{\tilde {\bf{M}}}}_{K \!\times n \times F}\! =\! \left\{{{\tilde m}_{k,i,j}}, \forall k \in {\cal{K}}, \forall i \! \in {\cal{N}}, \forall j \!\in {\cal{F}} \right\}$ is not integer. Then, the non-integer element should be branched. The whole procedure of branching can be divided into the following three steps:
%
\begin{itemize}
	\item \textit{Deriving Boolean Bound}
\end{itemize}

The integer variable ${{\tilde m}_{k',i',j'}},\forall k' \in {\cal{K}}, \forall i' \in {\cal{N}}, \forall j'\in {\cal{F}}$ is actually called Boolean variable. The near-optimal solution ${\bf{\tilde {\bf{M}}}}_{opt}$ can be initialized as ${\bf{\tilde {\bf{M}}}}_{opt}={\bf{\tilde {\bf{M}}}}_{r}$. Without loss of generality, in our non-linear mixed integer optimization problem to be solved, we use ${{\tilde m}_{k',i',j'}},\forall k' \in {\cal{K}}, \forall i' \in {\cal{N}}, \forall j'\in {\cal{F}}$ to denote the Boolean variable which does not satisfy the integer constraint and ${b_{k',i',j'}},\forall k' \in {\cal{K}}, \forall i' \in {\cal{N}}, \forall j'\in {\cal{F}}$ to denote its corresponding non-integer solution. Specifically, branch means that the relaxing problem \eqref{equ14} obtained from the original non-linear mixed integer problem should be further relaxed into two new sub-relaxing problems, which can be named sub-relaxing problem one and sub-relaxing problem two. Relaxing means adding two inequality constraints concerning Boolean variable ${{\tilde m}_{k',i',j'}},\forall k' \in {\cal{K}}, \forall i' \in {\cal{N}}, \forall j'\in {\cal{F}}$ to the relaxing problem \eqref{equ14} and keeping its original objective function and remaining constraints. To be specific, for the sub-relaxing problem one, the added inequality constraint can be expressed as $ {{\tilde m}_{k',i',j'}} \le \left\lfloor {{b_{k',i',j'}}} \right\rfloor , \forall k' \in {\cal{K}}, \forall i' \in {\cal{N}}, \forall j'\in {\cal{F}}$, while for the sub-relaxing problem two, the added inequality constraint is  ${{\tilde m}_{k',i',j'}} \ge \left\lfloor {{b_{k',i',j'}}} \right\rfloor + 1, \forall k' \in {\cal{K}}, \forall i' \in {\cal{N}}, \forall j'\in {\cal{F}}$, where $\left\lfloor  \cdot  \right\rfloor $ means rounding down.


It can be observed that, the added inequality constraint $ {{\tilde m}_{k',i',j'}} \le \left\lfloor {{b_{k',i',j'}}} \right\rfloor , \forall k' \in {\cal{K}}, \forall i' \in {\cal{N}}, \forall j'\in {\cal{F}}$ for the sub-relaxing problem one can combine with Eq. \eqref{11c} and merge into one equality constraint. Hence, the inequality constraint $ {{\tilde m}_{k',i',j'}} \le \left\lfloor {{b_{k',i',j'}}} \right\rfloor ,  \forall k' \in {\cal{K}}, \forall i' \in {\cal{N}}, \forall j'\in {\cal{F}}$ in the sub-relaxing problem one can be equivalently transformed into equality constraint ${{\tilde m}_{{k',i',j'}}} = 0, \forall k' \in {\cal{K}}, \forall i' \in {\cal{N}}, \forall j'\in {\cal{F}}$, which is called Boolean bound of 0 branch. Similarly, the added inequality constraint ${{\tilde m}_{k',i',j'}} \ge \left\lfloor {{b_{k',i',j'}}} \right\rfloor + 1, \forall k' \in {\cal{K}}, \forall i' \in {\cal{N}}, \forall j'\in {\cal{F}}$ for the sub-relaxing problem two can combine with Eq. \eqref{11c} and merge into one equality constraint. Therefore, the inequality constraint ${{\tilde m}_{k',i',j'}} \ge \left\lfloor {{b_{k',i',j'}}} \right\rfloor + 1, \forall k' \in {\cal{K}}, \forall i' \in {\cal{N}}, \forall j'\in {\cal{F}}$ in the sub-relaxing problem two can be equivalently transformed into the equality constraint ${{\tilde m}_{k',i',j'}} = 1, \forall k' \in {\cal{K}}, \forall i' \in {\cal{N}}, \forall j'\in {\cal{F}}$, which is called Boolean bound of 1 branch.
\begin{itemize}
	\item \textit{Formulating Two Sub-relaxing Problems}
\end{itemize}


Based on the obtained minimization relaxing problem \eqref{equ14} and Boolean bound of the 0-1 branches, we can easily get the sub-relaxing problem one and sub-relaxing problem two as follows:

Sub-relaxing problem one:
\begin{subequations}
	\begin{align}
		\mathop {\min }\limits_{{\bf{\tilde {\bf{M}}}}, {\bf{S}}} &\sum\limits_{j = 1}^F\left[ g\left( {{{\tilde m}_{k,i,j}},{s_{i,j}}} \right){\rm{ - }}{f\left( {{s_{i,j}}} \right)}\right]\\
		\text{s}\text{.t}\text{.}&\;\eqref{7b},\eqref{7c},\eqref{7d},\eqref{7e}\\
		&{{\tilde m}_{k',i',j'}} = 0,\forall k' \in {\cal{K}}, \forall i' \in {\cal{N}}, \forall j'\in {\cal{F}}\label{13c}
	\end{align}
    \label{equ_relax_one}
\end{subequations}
where constraint \eqref{13c} is the Boolean bound of 0 branch for the Boolean variable ${{\tilde m}_{k',i',j'}},\forall k' \in {\cal{K}}, \forall i' \in {\cal{N}}, \forall j'\in {\cal{F}}$.

Sub-relaxing problem two:
\begin{subequations}
	\begin{align}
		\mathop {\min }\limits_{{\bf{\tilde {\bf{M}}}}, {\bf{S}}} &\sum\limits_{j = 1}^F \left[g\left( {{{\tilde m}_{k,i,j}},{s_{i,j}}} \right){\rm{ - }}{f\left( {{s_{i,j}}} \right)}\right]\\
		\text{s}\text{.t}\text{.}&\;\eqref{7b},\eqref{7c},\eqref{7d},\eqref{7e}\\
		&{{\tilde m}_{k',i',j'}} = 1,\forall k' \in {\cal{K}}, \forall i' \in {\cal{N}}, \forall j'\in {\cal{F}}\label{14c}
	\end{align}
    \label{equ_relax_two}
\end{subequations}
where constraint \eqref{14c} is the Boolean bound of 1 branch for the Boolean variable ${{\tilde m}_{k',i',j'}},\forall k' \in {\cal{K}}, \forall i' \in {\cal{N}}, \forall j'\in {\cal{F}}$. 


\begin{itemize}
	\item \textit{Obtaining Near-Optimal Solution}
\end{itemize}

The sub-relaxing problem one formulated in Eq. \eqref{equ_relax_one} and the sub-relaxing problem two formulated in Eq. \eqref{equ_relax_two} can also be solved by GlobalSearch algorithm due to the fact that their forms are similar to the minimization relaxing problem \eqref{equ14}.

There are only two possible solutions for the two sub-relaxing problems expressed by Eq. \eqref{equ_relax_one} and Eq. \eqref{equ_relax_two}, i.e., infeasible solution and integer solution due to the Boolean bound of 0-1 branches. For the convenience of solving the problem, if the infeasible solution appears for the sub-relaxing problem, we can directly set it as positive infinity or negative infinity, which depends on whether the relaxing problem is a maximization problem or a minimization problem. Then, the near-optimal solution for one iteration of 0-1 branch and bound can be obtained by comparing the objective function values of two sub-relaxing problems to find the minimization branch and its corresponding solution, denoted as ${\bf{\tilde {\bf{M}}}}_{min}$ and ${\bf{S}}_{min}$, which are used to update the near-optimal solution ${\bf{\tilde {\bf{M}}}}_{opt}$ and ${\bf{S}}_{opt}$ for each iteration by letting ${\bf{\tilde {\bf{M}}}}_{opt}={\bf{\tilde {\bf{M}}}}_{min}$ and ${\bf{S}}_{opt}={\bf{S}}_{min}$. 

A new non-integer solution may appear in the updated solution ${\bf{\tilde {\bf{M}}}}_{opt}$ due to the 0-1 branch and bound. Therefore, if there is still one non-integer solution in the obtained ${\bf{\tilde {\bf{M}}}}_{opt}$, the 0-1 branch and bound should be executed continually based on the sub-relaxing problems formulated in Eq. \eqref{equ_relax_one} and Eq. \eqref{equ_relax_two}. To be specific, Boolean bound concerning the new non-integer solution should be added to the sub-relaxing problems \eqref{equ_relax_one} and \eqref{equ_relax_two}. Then, the near-optimal solution ${\bf{\tilde {\bf{M}}}}_{opt}$ and ${\bf{S}}_{opt}$ can be obtained by updating them as the near-optimal solution of the minimization branch of the current iteration of 0-1 branch and bound. Finally, the near-optimal solution of the original non-linear mixed integer optimization problem can be obtained until there is no non-integer solution in the updated near-optimal solution ${\bf{\tilde {\bf{M}}}}_{opt}$. The corresponding objective function value $Q_{min}$ can be obtained by substituting the near-optimal solution ${\bf{\tilde {\bf{M}}}}_{opt}$ and ${\bf{S}}_{opt}$ into the objective function of the optimization problem \eqref{equ14}. 

Generally speaking, to obtain the near-optimal integer solution of the considered optimization problem formulated in Eq. \eqref{equ14}, the maximum iteration to execute 0-1 branch and bound is $K \cdot n \cdot F$, which is actually the number of elements in the three-dimensional integer variable ${\bf{\tilde {\bf{M}}}}_{K \times n \times F} = \left\{{{\tilde m}_{k,i,j}}, \forall k \in {\cal{K}}, \forall i \in {\cal{N}}, \forall j \in {\cal{F}} \right\}$. Particularly, for each iteration of the 0-1 branch and bound, if there is more than one non-integer element, denoted as $N_n, {N_n} \ge 2$, in the updated solution ${\bf{\tilde {\bf{M}}}}_{opt}$, 0-1 branch and bound should be executed for each non-integer solution. Then, $2\cdot {N_n}$ branches are generated, which are compared to find the minimization branch and its corresponding solution ${\bf{\tilde {\bf{M}}}}_{min}$ and ${\bf{S}}_{min}$ to update the near-optimal solution as ${\bf{\tilde {\bf{M}}}}_{opt}={\bf{\tilde {\bf{M}}}}_{min}$, ${\bf{S}}_{opt}={\bf{S}}_{min}$. 

The solving procedure of our proposed near-optimal algorithm is summarized in Algorithm 1 named Near-Optimal Algorithm Based on Relaxation and Boolean bound.

\begin{algorithm}[!t]
	\label{alg1}
	\caption{Near-Optimal Algorithm Based on Relaxation and Boolean bound}
	\KwIn{$myfun1$: the objective function of optimzaton problem \eqref{equ14}; $myfun2$: non-linear constraints of optimization problem \eqref{equ14}; linear constraints of optimization problem \eqref{equ14}; ${\bf{\tilde {\bf{M}}}}_0$, $\bf{S}_0$: initial solution of ${\bf{\tilde {\bf{M}}}}$ and $\bf{S}$.}
	\KwOut{${\bf{\tilde {\bf{M}}}}_{opt}$, ${\bf{S}}_{opt}$ and $Q_{min}$.}
	\textbf{Initialize:} Obtain ${\bf{\tilde {\bf{M}}}}_{r}$, $\bf{S}_r$ and the objective function value $Q_{r}$ by solving relaxing problem \eqref{equ14}. Calculate the initial total number of non-integer elements in ${\bf{\tilde {\bf{M}}}}_{r}$, denoted as $N_n$, and obtain the position index ${{\bf{N}}_{in}}$.  Initialize the left coefficient matrix of equality constraints \eqref{13c} in Eq. \eqref{equ_relax_one} and \eqref{14c} in Eq. \eqref{equ_relax_two} as ${{\bf{A}}_{eq}} = {\bf{0}}$, and the right coefficient vector of equality constraints as ${{\bf{b}}_{eq}} = {\bf{0}}$.
	
	\eIf {$N_n$ = 0}
	{
		${\bf{\tilde {\bf{M}}}}_{opt}$ = ${\bf{\tilde {\bf{M}}}}_{r}$;\\
		${\bf{S}}_{opt}={\bf{S}}_r$;\\
		$Q_{min}$ = $Q_{r}$;
	}
	{
		
		Initialzie 	${\bf{\tilde {\bf{M}}}}_{opt}$ = ${\bf{\tilde {\bf{M}}}}_{r}$. Set the number of non-integer solution in ${\bf{\tilde {\bf{M}}}}_{opt}$ for each iteration of 0-1 branch and bound as $N_b$ = $N_n$ and the corresponding position index vector ${{\bf{N}}_i}$ = ${{\bf{N}}_{in}}$. 
		
		\For {$p = 1$ to $N_n$}
		{
			\For {$q = 1$ to $N_b$}
			{
				Update ${{\bf{A}}_{eq}}\left( {p,{{\bf{N}}_i}\left( {:,q} \right)} \right) = 1$; 
				
				Obtain the left coefficient matrix of equality constraints for 0 and 1 branch ${{\bf{A}}_{eq\_zero}} = {{\bf{A}}_{eq}}\left( {1:p,:} \right)$, ${{\bf{A}}_{eq\_one}} = {{\bf{A}}_{eq}}\left( {1:p,:} \right)$;
				
				Update ${{\bf{b}}_{eq}}\left( {p,:} \right) = 0$;
				
				Obtain the right coefficient vector of equality constraints for 0 branch ${{\bf{b}}_{eq\_zero}} = {{\bf{b}}_{eq}}\left( {1:p,:} \right)$;
				
				Update ${{\bf{b}}_{eq}}\left( {p,:} \right) = 1$;
				
				Obtain the right coefficient vector of equality constraints ${{\bf{b}}_{eq\_one}} = {{\bf{b}}_{eq}}\left( {1:p,:} \right)$;
				
				Solving the sub-relaxing problem one \eqref{equ_relax_one} and the sub-relaxing problem two \eqref{equ_relax_two} by utilizing the global optimal algorithm GlobalSearch.
			}
			Compare the objective function values of all 0-1 branches of $N_b$ non-integer variables, and find the minimization branch and obtain its corresponding solution, denoted as ${\bf{\tilde {\bf{M}}}}_{min}$ and ${\bf{S}}_{min}$;
			
			Update ${\bf{\tilde {\bf{M}}}}_{opt}$ by setting ${\bf{\tilde {\bf{M}}}}_{opt}={\bf{\tilde {\bf{M}}}}_{min}$, ${\bf{S}}_{opt}={\bf{S}}_{min}$;
			
			Update $N_b$ and ${{\bf{N}}_i}$ based on ${\bf{\tilde {\bf{M}}}}_{opt}$.
		}
		Obtain ${\bf{\tilde {\bf{M}}}}_{opt}$, ${\bf{S}}_{opt}$ and the objective function value $Q_{min}$ by substituting ${\bf{\tilde {\bf{M}}}}_{opt}$ and ${\bf{S}}_{opt}$ into the objective function $myfun1$ of optimization problem \eqref{equ14}.
	}
	
\end{algorithm}

\section{Suboptimal Solution Based on Greedy Method}

Considering that the near-optimal algorithm proposed is time-consuming, we propose a greedy algorithm with low computational complexity to obtain the suboptimal solution of original non-linear mixed 0-1 integer programming problem. The main idea of the greedy algorithm is to optimize variables ${\bf{\tilde {\bf{M}}}} = \left\{{{\tilde m}_{k,i,j}}, \forall k \in {\cal{K}}, \forall i \in {\cal{N}}, \forall j \in {\cal{F}} \right\}$ and  ${\bf{\bf{S}}} = \left\{{{\tilde s}_{i,j}}, \forall i \in {\cal{N}}, \forall j \in {\cal{F}} \right\}$ separately. The whole procedure is presented in the following two subsections.
\setlength{\belowdisplayskip}{2pt}
\vspace{-0.2cm}
\subsection{Caching Strategy Optimization with Encoding Parameters Fixed}
First, we optimize variable ${\bf{\tilde {\bf{M}}}}_{K \times n \times F} = \left\{{{\tilde m}_{k,i,j}}, \forall k \in {\cal{K}}, \forall i \in {\cal{N}}, \forall j \in {\cal{F}} \right\}$ with the encoding parameters fixed. The solution of variable ${\bf{\tilde {\bf{M}}}}_{K \times n \times F} = \left\{{{\tilde m}_{k,i,j}}, \forall k \in {\cal{K}}, \forall i \in {\cal{N}}, \forall j \in {\cal{F}} \right\}$ can be obtained by solving the optimization problem \eqref{equ10} under the condition of giving the variable ${\bf{\bf{S}}} = \left\{{{\tilde s}_{i,j}}, \forall i \in {\cal{N}}, \forall j \in {\cal{F}} \right\}$ a fixed value according to a given encoding rate. Actually, the solving process is the special case of Algorithm 1 with a fixed known ${{\bf{S}}}$, which greatly reduces the complexity of solving the optimization problem compared with a variable ${{\bf{S}}}$ since the dimension of the optimization variable reduces.

According to Algorithm 1, with a fixed known ${{\bf{S}}}$, the optimization problem \eqref{equ14}, \eqref{equ_relax_one} and \eqref{equ_relax_two} should be changed accordingly. Specifically, the minimization relaxing problem \eqref{equ14} can be reformulated as 
\begin{subequations}
	\begin{align}
		\mathop {\min }\limits_{{\bf{\tilde {\bf{M}}}}} &\sum\limits_{j = 1}^F g\left( {{{\tilde m}_{k,i,j}}} \right)\\
		\text{s}\text{.t}\text{.}&\;\eqref{7b},\eqref{7c},\eqref{7d},\eqref{11c}
	\end{align}
	\label{equ16}
\end{subequations}
\setlength{\abovedisplayskip}{7pt}
Based on the reformulated minimization relaxing problem \eqref{equ16}, the sub-relaxing problem one \eqref{equ_relax_one} and sub-relaxing problem two \eqref{equ_relax_two} can be reformulated as

\vspace{-0.1cm}
Sub-relaxing problem one:
\begin{subequations}
	\begin{align}
		\mathop {\min }\limits_{{\bf{\tilde {\bf{M}}}}}& \sum\limits_{j = 1}^F g\left( {{{\tilde m}_{k,i,j}}} \right)\\
		\text{s}\text{.t}\text{.}&\;\eqref{7b},\eqref{7c},\eqref{7d},\eqref{13c}
	\end{align}
	\label{equ17}
\end{subequations}
\vspace{-0.3cm}

Sub-relaxing problem two:
\begin{subequations}
	\begin{align}
		\mathop {\min }\limits_{{\bf{\tilde {\bf{M}}}}}& \sum\limits_{j = 1}^F g\left( {{{\tilde m}_{k,i,j}}} \right)\\
		\text{s}\text{.t}\text{.}&\;\eqref{7b},\eqref{7c},\eqref{7d},\eqref{14c}
	\end{align}
	\label{equ18}
\end{subequations}
\vspace{-0.9cm}
\subsection{Encoding Parameters Optimization Based on the Greedy Method}
Then, based on the caching strategy ${{\bf{\tilde M}}_g}$ obtained, the solution of variable ${\bf{\bf{S}}} = \left\{{{\tilde s}_{i,j}}, \forall i \in {\cal{N}}, \forall j \in {\cal{F}} \right\}$ can be obtained as ${{\bf{S}}_g}$ by gradually increasing the encoding rate for each video file one by one in a certain reasonable step. In this way, we can find the one which brings the maximum MOS value of the current iteration until not meeting the constraints in the optimization problem \eqref{equ14}. 

Finally, the suboptimal solution of the original non-linear mixed non-integer solution can be obtained as ${{\bf{\tilde M}}_g}$ and ${{\bf{S}}_g}$. The corresponding objective function value $Q_{min}$ can be obtained by substituting the suboptimal solution ${{\bf{\tilde M}}_g}$ and ${{\bf{S}}_g}$ into the objective function of the optimization problem \eqref{equ14}. The solving procedure for the greedy algorithm is summarized in Algorithm 2 named Suboptimal Algorithm Based on Greedy Method. 
\begin{algorithm}[!t]
	\label{alg2}
	\caption{Suboptimal Algorithm Based on Greedy Method}
	\KwIn{Caching strategy ${{\bf{\tilde M}}_g}$, which is obtained by utilizing Algorithm 1 with the known ${\bf{S}}$ based on the given encoding rate which satisfies constraint \eqref{7d} in optimization problem \eqref{equ14}.}
	\KwOut{${{\bf{\tilde M}}_g}$, ${\bf{S}}_{g}$ and $Q_{min}$.}
	\textbf{Initialize:} initialize ${\bf{S}}_g$ according to the minimum encoding rates; initialize $Q_{min}$ as $Q_{min}=Q_{g}$ which is calculated based on the obtained ${\bf{\tilde {\bf{M}}}}_{g}$ and ${\bf{S}}_{g}$; initialize increasing step of encoding rate as $\frac{1}{{{T_d} \cdot n}}$; initialize the searching times $N_1$ and $N_2$ as ${N_1} = \left( {{R_{\max }} - {R_{\min }}} \right) \cdot {T_d} \cdot N$, ${N_2} = F$, where $N$ is the number of video files whose encoding rate lies between ${R_{\min }}$ and ${R_{\max }}$. 
	
	\For {$p = 1$ to $N_1$}
	{
		\For {$q = 1$ to $N_2$}
		{
			Obtain all possible ${\bf{S}}_g$ which satisfy constraints in optimization problem \eqref{equ14} according to the increasing step $\frac{1}{{{T_d} \cdot n}}$.
			
			Obtain all possible $Q_{g}$ by substituting all possible ${\bf{S}}_g$, ${{\bf{\tilde M}}_g}$ into the objective function of optimization problem \eqref{equ14}. 
		}
		Compare all possible $Q_{g}$ and find the one which brings the most increments to $Q_{min}$, denoted as $Q_{gmin}$ and its corresponding encoding parameters, denoted as ${{\bf{S}}_{g\min }}$. 
		
		Update ${\bf{S}}_g$ by setting ${\bf{S}}_g={{\bf{S}}_{g\min }}$;
		
		Update $Q_{min}$ by setting $Q_{min}=Q_{gmin}$.
	}	
\end{algorithm}  
\section{Simulation Results}
In this section, we provide simulation results to highlight the performance of our algorithms. We consider a VoD downlink scenario in the cloud-edge network where the total number of video files is $F$, the total encoding packet of each video file is $n$, and the number of the EC servers is $K$. The total number of video requests received by SBSs during the whole delivery phase is $\psi $. The capacity of EC server is ${\Phi _k},\forall k \in K$. To reveal the performance of the proposed algorithms in scenarios with different configuration, the parameters $F,n,K,\psi ,{\Phi _k}$ can be adjusted in different figures. The tilt factor of the Zipf distribution is assumed as $\theta=0.8$ referring to relevant work \cite{litmm2018, liangtwc2017}. The duration of each video file is assumed as $T_d=2400s$ because the playback time of a video is generally 40min to 60min, here we choose 40min. The weighting coefficient of transmission latency is ${V_j} = 0.99,\forall j\in {\cal{F}}$ based on the MOS model in \cite{zhao2015tcsvt}. The transmission rate of the backhaul links is ${R_{bk}} = 10{\rm{Gbps/}}F$ according to the 3GPP specification \cite{3gpp}. To verify that encoding rates can be adjusted adaptively according to different caching capacity and security constraints, files of different encoding rates are considered. In particular, we consider three types of video files, the encoding rates of which satisfy different encoding rate ranges:

Type 1: The encoding rates of the 1st to the ($2/F$)-th video files vary from 0.3 Mbps to 0.7 Mbps; 

Type 2: The encoding rates of the (${2/F}+1$)-th to the (${2/F}+{4/F}$)-th video files vary from 1Mbps to 4 Mbps; 

Type 3: The encoding rates of the (${2/F}+{4/F}+1$)-th to the $F$-th video files vary from 4 Mbps to 8 Mbps. 

The detailed parameter settings are presented in Table \ref{tab:parameters}. Our proposed schemes and three existing algorithms are presented and analyzed as follows: 

\begin{itemize}
	\item \textit{EC-VE (Edge Caching and Video Encoding)}: Our proposed near-optimal edge caching and video encoding based algorithm that jointly optimizes video encoding parameters and edge caching strategy.   
\end{itemize}

\begin{itemize}
	\item \textit{Greedy EC-VE (Greedy Edge Caching and Video Encoding)}: Our proposed greedy edge caching and video encoding based algorithm with lower complexity, which optimizes video encoding parameters and edge caching strategy step by step.   
\end{itemize}

\begin{itemize}
	\item \textit{ECST (Edge Caching Secure Transmission)} in \cite{Gabry2016icc}: The edge caching based secure transmission scheme aims to minimize the backhaul links rate under secrecy constraints by utilizing edge caching strategy for general data transmission.    
\end{itemize}

\begin{itemize}
	\item \textit{EC-BF (Edge Caching and Beamforming)} in \cite{haoiot2020}: The secure transmission scheme aims to minimize delay in an edge cache-assisted millimeter-wave cloud radio access network by designing transmit and receive beamforming.  
\end{itemize}

\begin{itemize}
	\item \textit{EC-TM (Edge Caching and Trust Mechanism)} in \cite{xu2019itj}: The secure caching scheme is based on a trust mechanism and a Chinese remainder theorem-based privacy preservation protocol in heterogeneous networks for multi-homing users.  
\end{itemize}

Specifically, we validate the algorithm performance in terms of MOS, average MOS, transmission latency and the minimum caching capacity to ensure secure transmission with different simulation configurations. In order to show the performance difference under different video encoding rates and demonstrate the important impact of video encoding on system performance, three different fixed encoding rates are set for video files in the ECST scheme, EC-BF scheme and EC-TM scheme. In addition, ablation study about video encoding module and edge caching module is given in subsection \ref{sub1} to show the functions of the two modules and reveal their different effects on system performance. Finally, the computational complexity of the EC-VE algorithm, greedy EC-VE algorithm, the ECST algorithm, EC-BF algorithm and EC-TM algorithm is analyzed in subsection \ref{sub2}.  
\begin{table}[!t]
    \caption{VIDEO FILE PARAMETERS}\label{tab:parameters}
	\centering
	\scalebox{1}{
		\begin{tabular}{l l l}
			\hline
			Types & Encoding rate range & Video encoding quality parameters\\
			\hline
			Type 1 & 0.3Mbps-0.7Mbps & \begin{tabular}[c]{@{}l@{}}${C_1} = 0.23$, ${C_2} =  - 1.5$, \\ ${C_3} = 3.3$, ${C_4} = 2.5$ \end{tabular} \\
			Type 2 & 1Mbps-4Mbps & \begin{tabular}[c]{@{}l@{}}${C_1} = 0.0426$, ${C_2} =  - 0.4466$, \\ ${C_3} = 1.6369$, ${C_4} = 1.8415$ \end{tabular}\\
			Type 3 & 4Mbps-8Mbps & \begin{tabular}[c]{@{}l@{}}${C_1} = 0.0027$, ${C_2} =  - 0.0669$, \\ ${C_3} = 0.5842$, ${C_4} = 2.5248$ \end{tabular}\\
			\hline
	\end{tabular}}
\end{table}

\begin{figure}[!t]
	\centering
	\includegraphics[width=\linewidth]{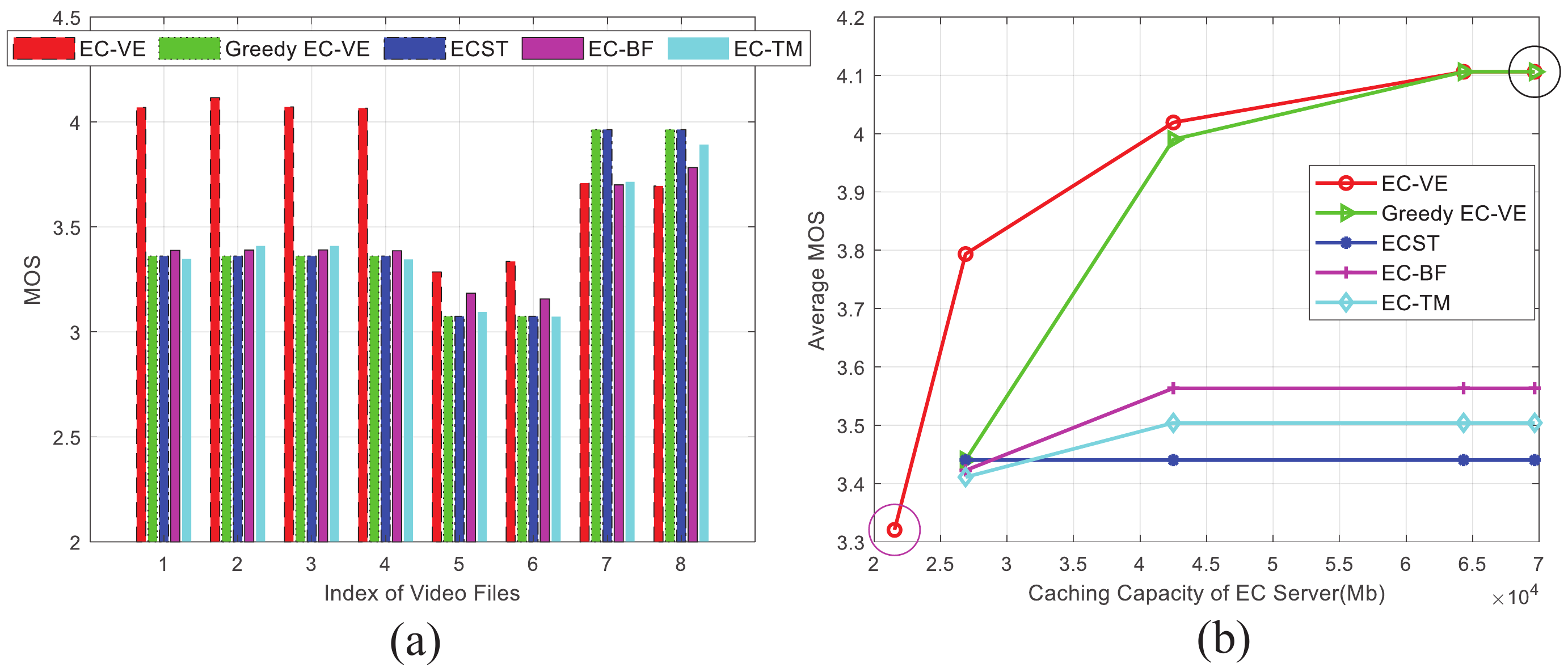}
	\caption{(a) MOS of each video file. (b) Average MOS vs. different caching capacity of EC server.}
	\label{fig1:mos}
\end{figure}
\vspace{-0.2cm}
\subsection{MOS of Each Video File}\label{suba}
\vspace{-0.1cm}
Fig. \ref{fig1:mos}. (a) shows the MOS value of each video file with $n=20$, $F=8$, $K=1$ and ${\Phi _k}=26880\text{Mb},\forall k \in {\cal{K}}$. The three fixed encoding rates of video files in the ECST scheme, EC-BF scheme and EC-TM scheme are ${{R_{\min }}}$, ${R_{\min }} + \left[ {{{\left( {{R_{\max }} - {R_{\min }}} \right)} \mathord{\left/{\vphantom {{\left( {{R_{\max }} - {R_{\min }}} \right)} {10}}} \right.\kern-\nulldelimiterspace} {10}}} \right]$ and ${R_{\min }} + \left[ {{{\left( {{R_{\max }} - {R_{\min }}} \right)} \mathord{\left/{\vphantom {{\left( {{R_{\max }} - {R_{\min }}} \right)} {20}}} \right.\kern-\nulldelimiterspace} {20}}} \right]$, respectively. Fig. \ref{fig1:mos}. (a) shows that, for each video file, the MOS performance of EC-VE algorithm is not always superior to the MOS performance of other schemes in spite of the obvious advantage in the average sense. The average MOS values of the EC-VE algorithm, Greedy EC-VE algorithm, ECST scheme, EC-BF scheme and EC-TM scheme are 3.7935, 3.4402 and 3.4402, 3.4225 and 3.4110, respectively. Our optimization goal is the whole performance advantage of all video files rather than the performance advantage for a single video file. Thus, there is no absolute superiority of MOS value for each video file in our proposed algorithm. Specifically, to obtain the whole performance advantage of the MOS value, the 1-st to 6-th video files, as depicted in Fig. \ref{fig1:mos}. (a), which can get higher QoE with lower encoding rate, are more likely to be distributed with caching capacity to improve QoE on the condition of limited caching capacity, while the 7-th and 8-th video files, which request higher encoding rate, are distributed with less caching capacity.

Each video file has own caching strategy and encoding parameter. The number and the size of encoding packets cached on the EC server vary for different video files. Therefore, the video encoding quality and transmission latency of each video file are different for the five schemes. This results in different MOS values in the case of the same caching capacity of the EC server. Furthermore, the MOS values of the Greedy EC-VE algorithm and ECST scheme are the same for each video file. This is because each video file of both schemes is encoded into the minimum encoding rate. Meanwhile, the caching strategy for each video file is the same when the caching capacity is ${\Phi _k}=26880\text{Mb},\forall k \in {\cal{K}}$. This value of caching capacity allows all encoding packets of each video file to be cached on the EC server at the minimum encoding rate. Besides, the MOS values of the EC-BF scheme and the EC-TM scheme are lower than that of the other schemes. Because for the EC-BF scheme and EC-TM scheme, the secure transmission rate for each video file is lower than that of other schemes. Moreover, for the EC-TM scheme, the caching capacity for improving MOS is more scarce compared with other schemes due to the fact that the encrypted video files are larger and occupy more caching capacity.
\vspace{-0.2cm}
\subsection{Average MOS with Different Caching Capacity of EC Server}\label{subb}  

In Fig. \ref{fig1:mos}. (b), the average MOS value versus different caching capacity of EC server with $n=20$, $F=8$, $K=1$ is presented. The three fixed encoding rates of video files in the ECST scheme, EC-BF scheme and EC-TM scheme are ${{R_{\min }}}$, ${R_{\min }} + \left[ {{{\left( {{R_{\max }} - {R_{\min }}} \right)} \mathord{\left/{\vphantom {{\left( {{R_{\max }} - {R_{\min }}} \right)} {10}}} \right.\kern-\nulldelimiterspace} {10}}} \right]$ and ${R_{\min }} + \left[ {{{\left( {{R_{\max }} - {R_{\min }}} \right)} \mathord{\left/{\vphantom {{\left( {{R_{\max }} - {R_{\min }}} \right)} {20}}} \right.\kern-\nulldelimiterspace} {20}}} \right]$, respectively. As shown in Fig. \ref{fig1:mos}. (b), the average MOS values of the EC-VE algorithm and the Greedy EC-VE algorithm increase with the increasing caching capacity of the EC server. Both of them eventually tend to the same maximum value, which is determined by the maximum encoding rate of each video file and the optimal caching strategy. At this time, the caching capacity of EC server is sufficient to cache all encoding packets. The EC-VE algorithm also achieves a considerable performance gain compared with the Greedy EC-VE algorithm. The minimum average MOS value of the proposed EC-VE algorithm is obtained when the caching capacity of the EC server is just enough to cache the minimum number of encoding packets to satisfy the security constraints. However, the Greedy EC-VE algorithm cannot find the feasible solution that satisfies the security constraints at this point, marked by the purple circle in Fig. \ref{fig1:mos}. (b). As the caching capacity of the EC server increases, the gap of the average MOS values is initially large, then gradually narrows. This is because large caching capacity can provide more choices of caching strategy and encoding rate to bring better performance on video encoding quality and transmission latency.

Furthermore, the average MOS value of the ECST scheme, which is always worse than the average MOS value of the proposed EC-VE algorithm and Greedy EC-VE algorithm, remains unchanged with the increasing caching capacity. This scenario is reasonable since the minimum encoding rate is adopted in the ECST scheme. At this time, the caching strategy is the same regardless of the caching capacity because its aim is to minimize the backhaul links rate. Thus, the video encoding packets cached on the EC server will be as many as possible if the caching capacity is large enough and all encoding packets are cached on the EC server when each video file is encoded into its minimum encoding rate. 

Besides, the average MOS values of the EC-BF scheme and the EC-TM scheme, which are much worse than that of the proposed EC-VE algorithm and Greedy EC-VE algorithm, increase with the increasing caching capacity initially and gradually remain stable. In the beginning, less encoding packets are cached on EC servers to ensure security due to the limited caching capacity. With the increasing caching capacity, all encoding packets can be cached on EC servers to ensure security and reduce transmission latency at the same time. However, the MOS values of the EC-BF scheme and EC-TM scheme keep stable when the caching capacity continues to increase as a result of the fixed encoding rates. The gap of the EC-BF scheme and the EC-TM scheme results from different encoding rate and the effective utilization rate of caching capacity.

The Greedy EC-VE algorithm, the ECST scheme, the EC-BF scheme and the EC-TM scheme cannot achieve secure transmission when the caching capacity value is very small, i.e., at the first point marked by the purple circle in Fig. \ref{fig1:mos}. (b) with caching capacity ${\Phi _k}=21540\text{Mb},\forall k \in {\cal{K}}$, while the proposed EC-VE algorithm can ensure secure transmission and obtain a feasible average MOS value by adjusting caching strategy and encoding parameters at the same time. The observations above demonstrate the benefits of the proposed algorithm, which comprehensively considers the interaction of caching strategy and encoding parameters, and adjusts both of them simultaneously to adapt to different caching capacity.

\begin{table}[!t]\scriptsize
	\caption{MOS OF EACH VIDEO FILE UNDER DIFFERENT CACHING CAPACITY}\label{tab:simulation}
	\centering
	\scalebox{0.95}{
		\begin{tabular}{l m{0.4cm} m{0.4cm} m{0.4cm} m{0.4cm} m{0.4cm} m{0.4cm} m{0.4cm} m{0.4cm}}
			\hline
			\multicolumn{9}{c}{MOS of Each Video File under the Smallest Caching Capacity}\\
			\hline
			\begin{tabular}[c]{@{}l@{}}Index of\\ Video File\end{tabular} & 1 & 2 & 3 & 4 & 5 & 6 & 7 & 8\\
			\begin{tabular}[c]{@{}l@{}}Encoding Rate\\ (Mbps)\end{tabular} & 0.3 & 0.3 & 0.3 & 0.3 & 1 & 1 & 4 & 4\\
			\begin{tabular}[c]{@{}l@{}}Number of\\Cached Packets\end{tabular}& 15 & 19 & 17 & 14 & 18 & 18 & 16 & 15\\
			MOS & 3.302 & 3.303 & 3.303 & 3.300 & 2.984 & 2.966 & 2.708 & 2.695\\
			\hline
			\multicolumn{9}{c}{MOS of Each Video File under the Largest Caching Capacity}\\
			\hline
			\begin{tabular}[c]{@{}l@{}}Index of\\Video File\end{tabular} & 1 & 2 & 3 & 4 & 5 & 6 & 7 & 8\\
			\begin{tabular}[c]{@{}l@{}}Encoding Rate\\(Mbps)\end{tabular} & 0.7 & 0.7 & 0.7 & 0.7 & 4 & 4 & 8 & 8\\
			\begin{tabular}[c]{@{}l@{}}Number of \\Cached Packets\end{tabular}& 20 & 20 & 20 & 20 & 20 & 20 & 20 & 20\\
			MOS & 4.116 & 4.116 & 4.116 & 4.116 & 3.932 & 3.932 & 4.261 & 4.261\\
			\hline
	\end{tabular}}
\end{table}

To show the changes of encoding parameters and caching strategy under different caching capacity, encoding rates and the number of cached packets for two selected points in Fig. \ref{fig1:mos}. (b) are given in Table \ref{tab:simulation}. At the first point marked by the purple circle in Fig. \ref{fig1:mos}. (b), the minimum average MOS value of the proposed EC-VE algorithm is obtained. At the last point marked by the black circle in Fig. \ref{fig1:mos}. (b), the same maximum average MOS values of the proposed EC-VE algorithm and Greedy EC-VE algorithm are obtained. Table \ref{tab:simulation} gives the MOS value of each video file at the first point when the EC server has the smallest caching capacity ${\Phi _k}=21540\text{Mb},\forall k \in {\cal{K}}$ and at the last point when the EC server has the largest caching capacity ${\Phi _k}=69660\text{Mb},\forall k \in {\cal{K}}$. The minimum MOS of each video file is obtained at the first point. In this situation, all video files in the EC-VE algorithm are encoded into the minimum encoding rates, and the minimum number of encoding packets are cached on the EC server to satisfy the security constraints due to the limited caching capacity. Meanwhile, the minimum number of cached packets varies for different video files, because different video files have a different number of requests. At the last point, the maximum MOS of each video file is obtained, because all video files of both algorithms are encoded into the maximum encoding rates and all encoding packets are cached on the EC server due to the sufficient caching capacity.
\vspace{-0.2cm}
\subsection{Average MOS with Different Parameters of Video Files}\label{subc}

\begin{figure}[!t]
	\centering
	\includegraphics[width=\linewidth]{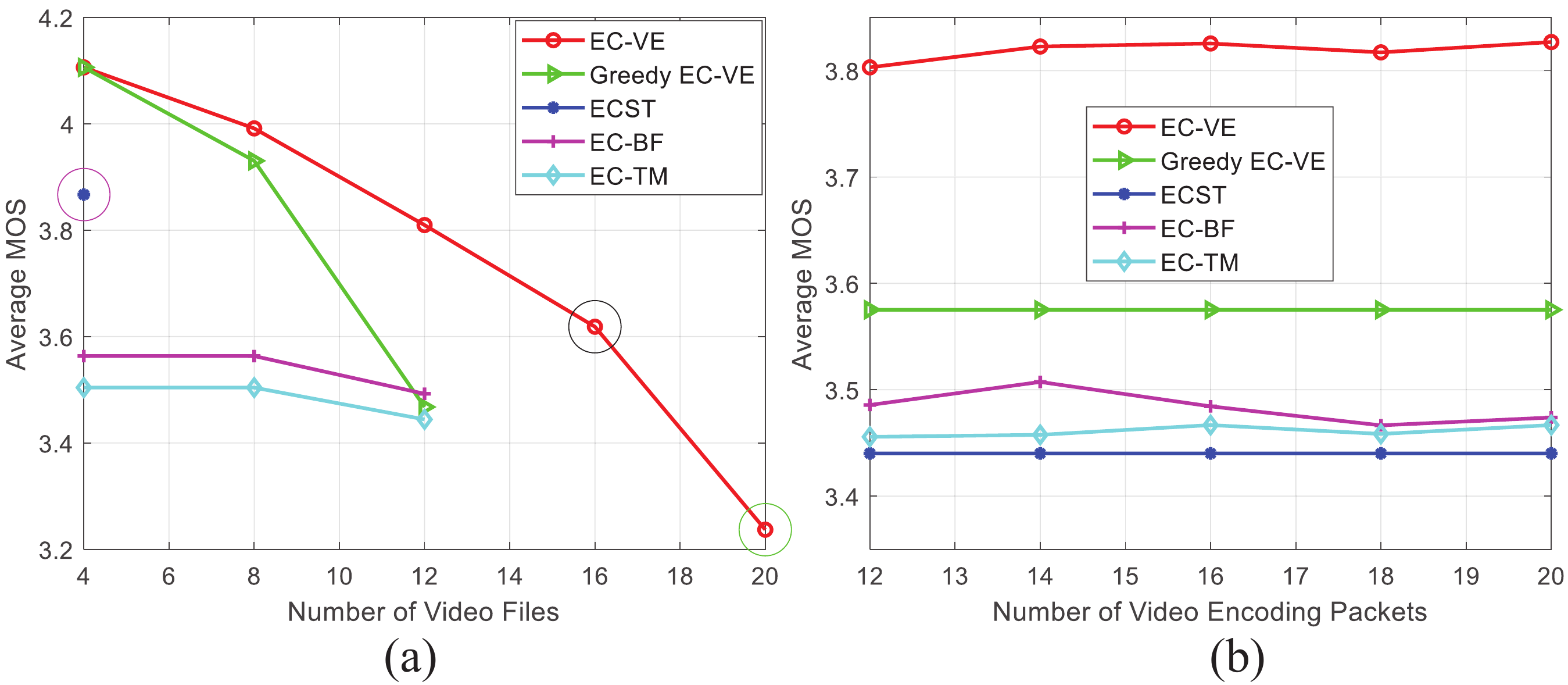}
	\caption{(a) Average MOS vs. different number of video files. (b) Average MOS vs. different number of video encoding packets.}
	\label{fig56:mos}
\end{figure}

Fig. \ref{fig56:mos}. (a) shows the average MOS value versus different number of video files with $n=10$, $K=1$ and ${\Phi _k}=39000\text{Mb},\forall k \in {\cal{K}}$. The three fixed encoding rates of video files in the ECST scheme, EC-BF scheme and EC-TM scheme are ${R_{\min }} + \left[ {{{\left( {{R_{\max }} - {R_{\min }}} \right)} \mathord{\left/{\vphantom {{\left( {{R_{\max }} - {R_{\min }}} \right)} {2}}} \right.\kern-\nulldelimiterspace} {2}}} \right]$, ${R_{\min }} + \left[ {{{\left( {{R_{\max }} - {R_{\min }}} \right)} \mathord{\left/{\vphantom {{\left( {{R_{\max }} - {R_{\min }}} \right)} {10}}} \right.\kern-\nulldelimiterspace} {10}}} \right]$ and ${R_{\min }} + \left[ {{{\left( {{R_{\max }} - {R_{\min }}} \right)} \mathord{\left/{\vphantom {{\left( {{R_{\max }} - {R_{\min }}} \right)} {20}}} \right.\kern-\nulldelimiterspace} {20}}} \right]$, respectively. From Fig. \ref{fig56:mos}. (a), the average MOS value decreases with the increasing number of video files for the proposed EC-VE algorithm and the Greedy EC-VE algorithm. The gap between them increases with the increasing video files when the number of video files is less than or equal to $F=12$. However, the Greedy EC-VE algorithm has no solution as the number of video files continues to increase to $F=16$ and $F=20$, marked by the black and cyan circles in Fig. \ref{fig56:mos}. (a), while the EC-VE algorithm still has the feasible and effective solution. Specifically, the minimum average MOS value is obtained at the last point marked by the cyan circle. At this time, all video files are encoded into the minimum encoding rates, and the minimum number of encoding packets are cached on the EC server. Besides, the average MOS values of the EC-BF scheme and the EC-TM scheme, which are worse than our proposed schemes, keep stable when the number of video files is less than or equal to $F=8$. Then, the MOS values of them decrease when the video files increase to $F=12$.

The reasons contributing to these observations are that when there are four video files, all of them are encoded into the maximum encoding rates in the proposed near-optimal and greedy algorithm, and encoded into the fixed encoding rates in the EC-BF scheme and the EC-TM scheme. Meanwhile, all encoding packets are cached on the EC server due to the sufficient caching capacity. This brings the best MOS values for each scheme as shown by the first points. As the number of video files increases, only a part of the encoding packets can be cached on the EC server due to the limited caching capacity. This cannot allow all video files to be encoded into the maximum encoding rates. In this situation, the caching strategy and encoding parameters of the proposed two algorithms are different. For the EC-VE algorithm, the near-optimal average MOS value can be obtained by jointly optimizing the caching strategy and encoding parameters flexibly, while the Greedy EC-VE algorithm achieves worse MOS values or even an infeasible solution. The reason is that it optimizes the caching strategy and encoding parameters step by step. Thus, it cannot satisfy the complex and strict security constraints when there are many video files. For the EC-BF scheme and the EC-TM scheme, they cannot ensure security when the video files increase to $F=12$ because of the fixed encoding rates and limited caching capacity. This phenomenon further demonstrates the performance advantage of the near-optimal EC-VE algorithm. 

Furthermore, it is interesting to find that there is only one feasible point marked by the purple circle in Fig. \ref{fig56:mos}. (a) for the ECST scheme when there are four video files in the considered scenario, and the performance of its average MOS value is much worse than the proposed algorithms. This is because the ECST scheme cannot satisfy the security constraints by adjusting its caching strategy when there are many video files and the caching capacity is limited. Meanwhile, the simple caching strategy of the ECST scheme demands that the number of encoding packets cached in different EC servers must be the same for the same video file. 

In Fig. \ref{fig56:mos}. (b), the average MOS value versus different number of video encoding packets with $F=8$, $K=1$ and ${\Phi _k}=28000\text{Mb},\forall k \in {\cal{K}}$ is presented. The three fixed encoding rates of video files in the ECST scheme, EC-BF scheme and EC-TM scheme are ${{R_{\min }}}$, ${R_{\min }} + \left[ {{{\left( {{R_{\max }} - {R_{\min }}} \right)} \mathord{\left/{\vphantom {{\left( {{R_{\max }} - {R_{\min }}} \right)} {10}}} \right.\kern-\nulldelimiterspace} {10}}} \right]$ and ${R_{\min }} + \left[ {{{\left( {{R_{\max }} - {R_{\min }}} \right)} \mathord{\left/{\vphantom {{\left( {{R_{\max }} - {R_{\min }}} \right)} {20}}} \right.\kern-\nulldelimiterspace} {20}}} \right]$, respectively. As shown in Fig. \ref{fig56:mos}. (b), the average MOS values of the proposed EC-VE algorithm, Greedy EC-VE algorithm, the ECST scheme, EC-BF scheme and EC-TM scheme are almost stable with the increasing number of encoding packets. The MOS performance of the proposed near-optimal EC-VE algorithm fluctuates slightly when the number of encoding packets changes. Meanwhile, it is significantly superior to the Greedy EC-VE algorithm, the ECST scheme, the EC-BF scheme and EC-TM scheme. Specifically, the MOS performance increases slowly at first, then decreases when the encoding packets increase to $n=18$, and finally increases to the maximum value when the number of encoding packets is $n=20$. The fluctuations result from the caching strategy and the encoding parameters which vary slightly with the changing total number of encoding packets, when the caching capacity and the number of video files remain unchanged. A tradeoff exists between the total number of encoding packets and the size of each encoding packet, which results in the falling fluctuation when the number of encoding packets is $n=18$. In this situation, the proposed near-optimal EC-VE algorithm will wisely consider this balance and achieve the near-optimal results by joint adjustment of caching strategy and encoding parameters. This also brings the reasonable fluctuations. 

It can also be observed that for the EC-BF scheme and EC-TM scheme, there are also fluctuations, especially for the EC-BF scheme. This is because the caching strategy and secure transmission rate adjusts within the available caching capacity when the number of encoding packets changes, which brings difference of transmission latency. As for the Greedy EC-VE algorithm, its performance of the average MOS value is still better than that of the comparable ECST scheme, while the average MOS values of both schemes almost remain stable when the number of encoding packets changes. This is because both of them optimize the caching strategy with the fixed encoding parameters, which results in the same caching strategy when the caching capacity of the EC server and the number of video files keep unchanged in spite of the changing number of total encoding packets.
\vspace{-0.3cm}
\subsection{Average MOS with Different Number of EC Servers}\label{subd}

\begin{figure}[!t]
	\centering
	\includegraphics[width=\linewidth]{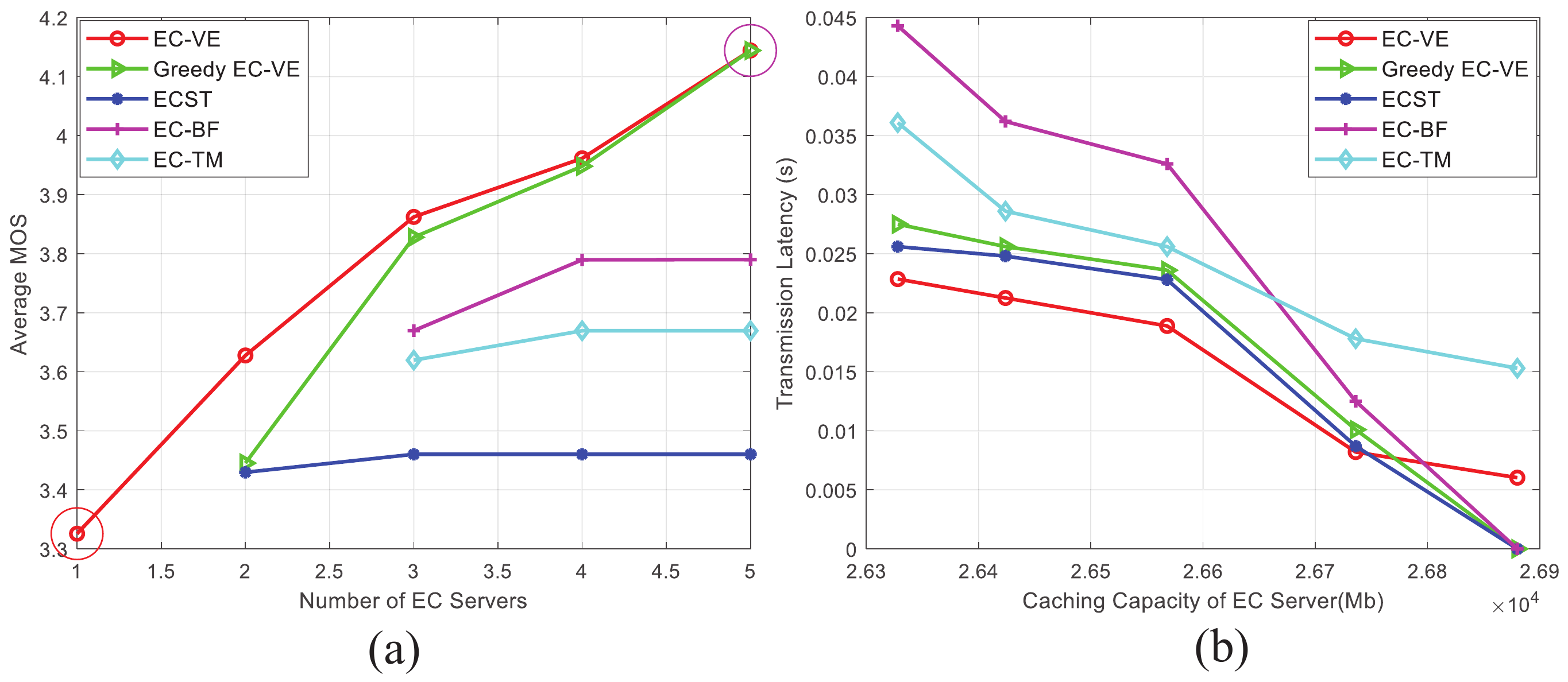}
	\caption{(a) Average MOS vs. different number of EC servers. (b) Transmission latency vs. different caching capacity of EC server.}
	\label{figure1}
\end{figure}

Fig. \ref{figure1}. (a) shows the average MOS value versus different number of EC servers with $n = 10,F = 12$. The three fixed encoding rates of video files in the ECST scheme, EC-BF scheme and EC-TM scheme are ${R_{\min }} + \left[ {{{\left( {{R_{\max }} - {R_{\min }}} \right)} \mathord{\left/{\vphantom {{\left( {{R_{\max }} - {R_{\min }}} \right)} {10}}} \right.\kern-\nulldelimiterspace} {10}}} \right]$, ${R_{\min }} + \left[ {{{\left( {{R_{\max }} - {R_{\min }}} \right)} \mathord{\left/{\vphantom {{\left( {{R_{\max }} - {R_{\min }}} \right)} {2}}} \right.\kern-\nulldelimiterspace} {2}}} \right]$ and ${R_{\min }} + \left[ {{{\left( {{R_{\max }} - {R_{\min }}} \right)} \mathord{\left/{\vphantom {{\left( {{R_{\max }} - {R_{\min }}} \right)} {5}}} \right.\kern-\nulldelimiterspace} {5}}} \right]$, respectively. Considering more general scenarios, EC servers are assumed to have different caching capacity. To be specific, when the number of EC servers is $K>1$, the caching capacity values of EC servers are $[21271\;  18272]{\rm{Mb}}$, $[21271\;18272\;13017]{\rm{Mb}}$, $[21271\;18272\;13017\;22000]{\rm{Mb}}$, $[21271\;18272\;13017\;22000\;22224]{\rm{Mb}}$, respectively. As shown in Fig. \ref{figure1}. (a), the average MOS values of the EC-VE algorithm and Greedy EC-VE algorithm increase with the increasing number of EC servers. They reach to the same maximum value when the total caching capacity of all EC servers is sufficient. In this case, encoding packets of all video files are cached on EC servers in the maximum encoding rates, just as the last point shows marked by purple circle in Fig. \ref{figure1}. (a). The performance of the EC-VE scheme is significantly superior to other schemes.

Besides, the average MOS values of the ECST scheme, the EC-BF scheme and the EC-TM scheme initially rise with the increasing number of EC servers, then remain stable. The reason is that less encoding packets are cached on EC servers due to the limited total caching capacity of EC servers. As the number of EC servers increases, the provided caching capacity is sufficient to cache more encoding packets on EC servers until all encoding packets are cached on EC servers in the fixed encoding rates. This leads to the stable average MOS of the three schemes when the number of EC servers increase to four. It is also interesting to find that there exist infeasible solutions for each scheme except for the EC-VE scheme. This is because the joint optimization of video encoding and caching strategy can wisely utilize the limited caching capacity provided by one EC server to improve QoE and ensure security by flexibly adjusting encoding parameters and caching strategy while other schemes cannot, just as the first point shows marked by red circle in Fig. \ref{figure1}. (a). 
\vspace{-0.3cm}
\subsection{Transmission Latency with Different Caching Capacity of EC Server}\label{sube}

In Fig. \ref{figure1}. (b), the transmission latency versus different caching capacity of EC server with $n=20$, $F=8$, $K=1$ is presented. To solely show the performance of transmission latency, the fixed encoding rates of video files in the ECST scheme, EC-BF scheme and EC-TM scheme are set as the same encoding rate, i.e., ${{R_{\min }}}$. As shown in Fig. \ref{figure1}. (b), the transmission latency of the EC-VE scheme, Greedy EC-VE scheme, ECST scheme, EC-BF scheme and EC-TM scheme decreases with the increasing caching capacity of EC server. The transmission latency of the Greedy EC-VE scheme, ECST scheme and EC-TM scheme decreases to zero when the caching capacity is ${\Phi _k}=26880\text{Mb},\forall k \in {\cal{K}}$ because all encoding packets can be cached on EC server at the fixed minimum encoding rates due to the sufficient caching capacity. In general, the transmission latency of the EC-VE scheme is lower than the ECST scheme, which is lower than other schemes. This phenomenon strongly proves the performance advantage of the proposed EC-VE scheme in reducing transmission latency. It can find the tradeoff between video encoding quality and transmission latency by adjusting the cached encoding packets sizes when caching capacity is limited, which leads to its non-zero value at the last point in Fig. \ref{figure1}. (b). Besides, the ECST scheme outperforms the Greedy EC-VE scheme. This is because the goal of the ECST scheme is to minimize the backhaul links rate. It aims to cache as many encoding packets as possible on EC servers with available caching capacity. Besides, the transmission latency of the EC-TM scheme does not decrease to zero because more caching capacity is needed to cache the encrypted video files.
\vspace{-0.1cm}
\subsection{Minimum Caching Capacity with Different Number of Requests of Users}\label{subf}
\begin{figure}[!t]
	\centering
	\includegraphics[width=80mm]{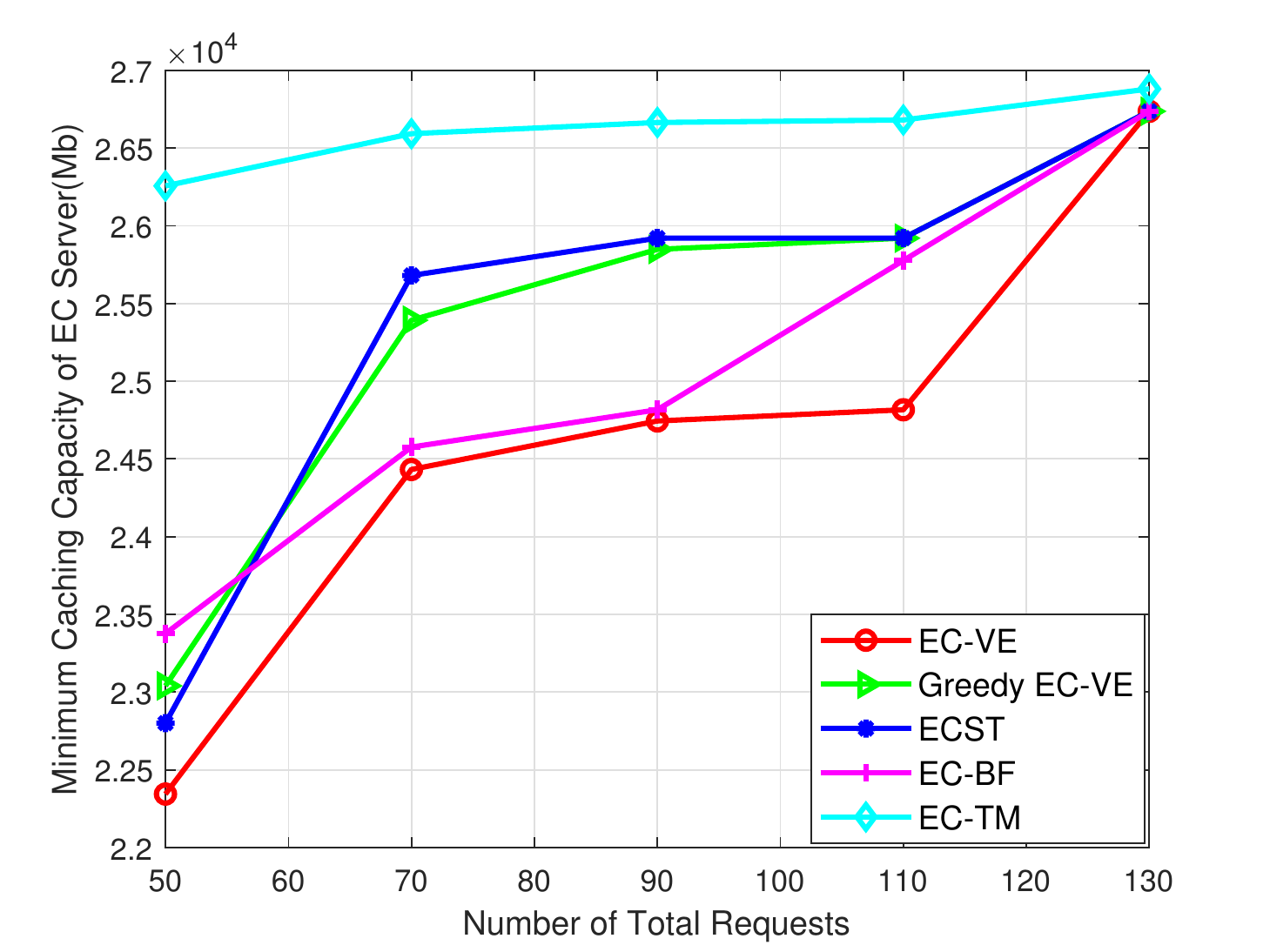}
	\caption{Minimum caching capacity vs. different number of total requests.}
	\label{fig6:mos}
\end{figure}
\vspace{-0.1cm}
Fig. \ref{fig6:mos} shows the minimum caching capacity of the EC server versus different number of total requests with $n=10$, $F=8$, $K=1$. Because the minimum caching capacity is required, the fixed encoding rates of video files in the ECST scheme, EC-BF scheme and EC-TM scheme are set as the minimum encoding rate, i.e., ${{R_{\min }}}$. The minimum caching capacity of the five schemes increases with the increasing number of total requests and eventually tends to the same value when the number of total requests reaches $\psi=130$. This is because more caching capacity is needed to cache more video encoding packets to ensure the requirements of video encoding quality, transmission latency and security constraints as the number of total requests increases. Furthermore, Fig. \ref{fig6:mos} shows that the minimum caching capacity of the proposed EC-VE algorithm is much smaller than that of the Greedy EC-VE algorithm, the ECST scheme and the EC-TM scheme. This phenomenon shows that the proposed EC-VE algorithm still has relatively good performance even with limited caching resources. Meanwhile, it can spare more caching resources to achieve better video encoding quality and low latency when satisfying the security constraints. 

It is interesting to find that the gap of the minimum caching capacity between EC-VE algorithm and the EC-BF scheme is relatively small when the number of total requests is $\psi=70$ and $\psi=90$. This reflects the advantage of the rate adaptive scheme EC-BF compared with other schemes. Meanwhile, the minimum caching capacity of the EC-TM scheme is much larger than other schemes because encrypted video files require more caching capacity under the same conditions. The minimum caching capacity of other four schemes except EC-TM scheme finally reaches the same value as the number of total requests increases to $\psi=130$, because more requests mean that more video encoding packets should be cached on the EC server for any scheme. As a result, almost all video encoding packets are cached on the EC server at the minimum encoding rates to satisfy the security constraints and guarantee the video encoding quality and latency requirements when the number of total requests reaches $\psi=130$. 
\vspace{-0.2cm}
\subsection{Ablation Study about Video Encoding and Edge Caching}\label{sub1}
In this subsection, the impact of video encoding module and edge caching module on the performance of MOS, average MOS and the minimum caching capacity to ensure secure transmission is analyzed. For comparison, in addition to our proposed near-optimal EC-VE algorithm, the other two algorithms which are solely based on video encoding or edge caching are considered and presented:

\begin{itemize}
	\item \textit{EC (Edge Caching)}: Edge caching based algorithm which solely utilizes edge caching to improve video encoding quality and reduce latency under secure transmission for the backhaul links.   
\end{itemize}

\begin{itemize}
	\item \textit{VE (Video Encoding)}: Video encoding based algorithm which solely utilizes video encoding to improve video encoding quality and reduce latency under secure transmission for the backhaul links.   
\end{itemize}

\begin{figure}[!t]
	\centering
	\includegraphics[width=80mm]{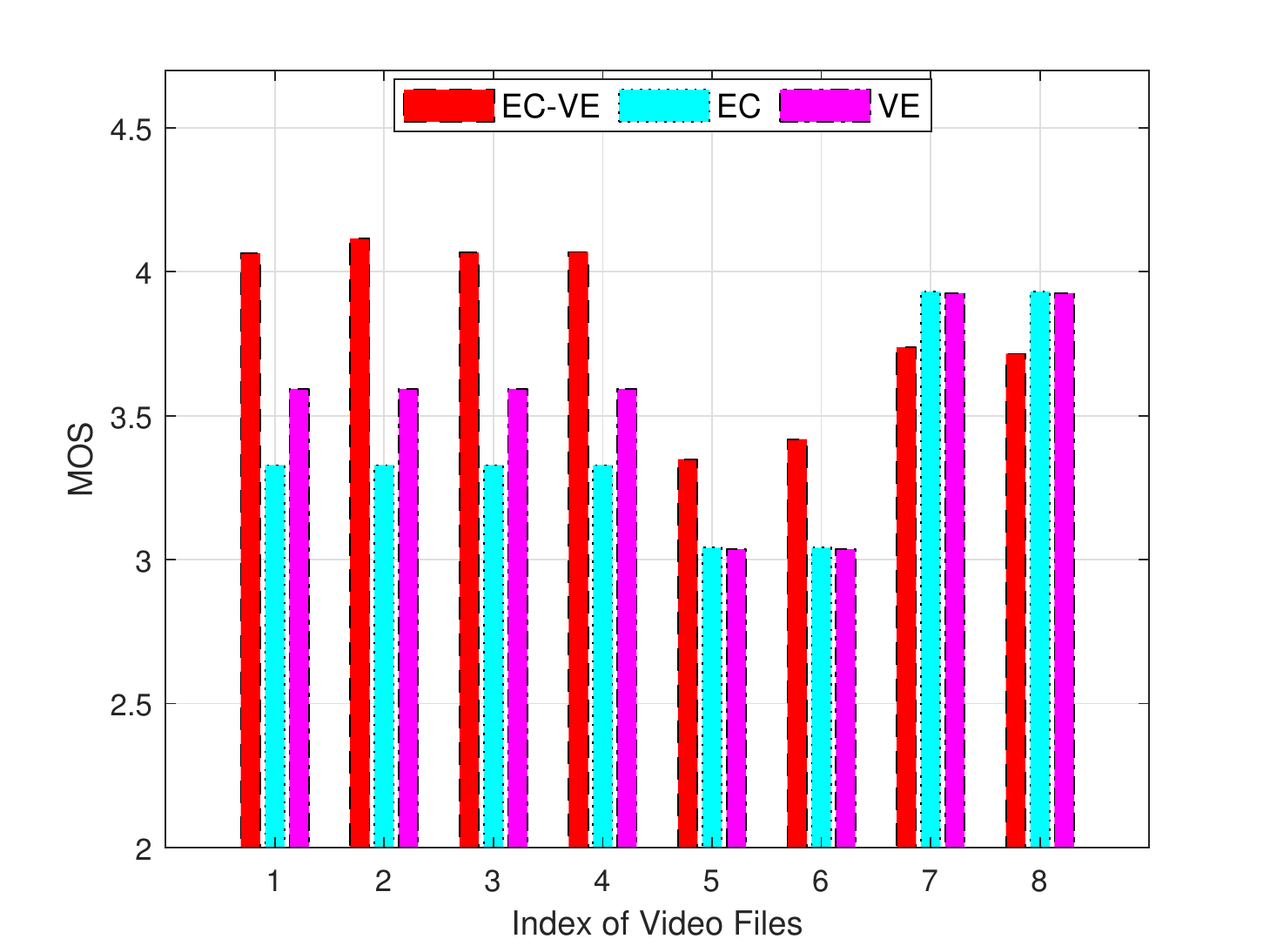}
	\caption{MOS of each video file.}
	\label{fig9:mos}
\end{figure}

Fig. \ref{fig9:mos} shows the MOS value of each video file with $F=8$, $n=18$, $K=1$ and ${\Phi _k}=28000\text{Mb},\forall k \in {\cal{K}}$. From Fig. \ref{fig9:mos}, for each video file, the MOS values of the EC-VE algorithm, EC algorithm and VE algorithm are different. The average MOS values of them are 3.8172, 3.4075 and 3.5371, respectively. Although the EC-VE algorithm outperforms the other two in the average sense, for each video file, it is not always better. This is because the EC-VE scheme can obtain the whole performance superiority by jointly adjusting edge caching strategy and video encoding parameters, while the EC scheme and VE scheme can improve MOS for certain files by only utilizing edge caching or video encoding strategy. As shown in Fig. \ref{fig9:mos}, the 1-st to 6-th video files are more likely to be allocated more caching capacity to achieve higher QoE with relatively low encoding rate, while the 7-th and 8-th video files are allocated less caching capacity. This is because the QoE requirements of the 7-th and 8-th video files with high encoding rates demand more caching capacity. Furthermore, the MOS performance of the VE scheme performs better than that of the EC scheme. This demonstrates the advantage of video encoding strategy with limited caching capacity. 

\begin{figure}[!t]
	\centering
	\includegraphics[width=\linewidth]{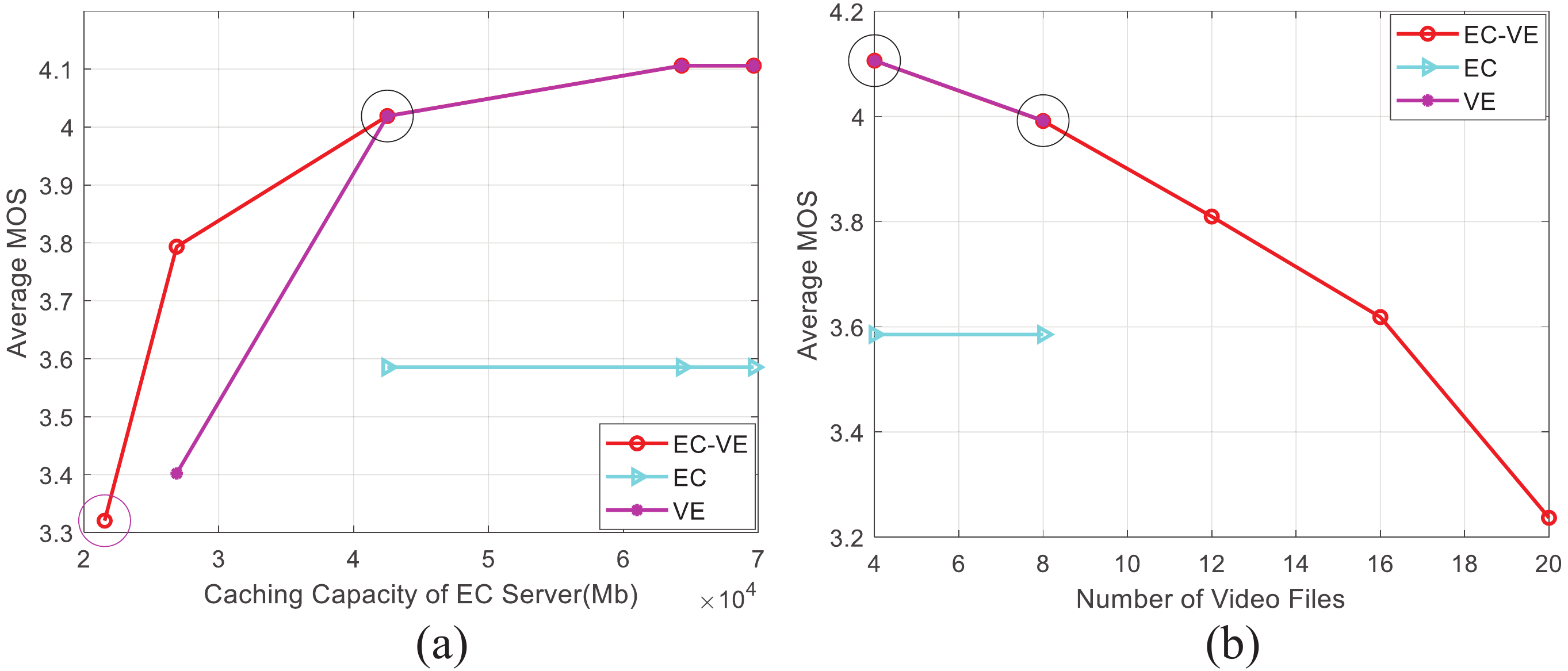}
	\caption{(a) Average MOS vs. different caching capacity of EC server. (b) Average MOS vs. different number of video files.}
	\label{fig1011:mos}
\end{figure}

Fig. \ref{fig1011:mos}. (a) shows the average MOS versus different caching capacity of EC server with $F=8$, $n=20$ and $K=1$. As shown in Fig. \ref{fig1011:mos}. (a), the average MOS values of the EC-VE scheme and VE scheme rise with increasing caching capacity, while the average MOS value of the EC scheme keeps stable when caching capacity increases. This is because for the EC scheme, all packets are cached on EC server with the fixed minimum encoding rates which cannot be adjusted when the caching capacity changes, while the EC-VE scheme and VE scheme can obtain higher QoE by encoding video files into higher encoding rates when more caching capacity are provided. Furthermore, it is interesting to find that the VE scheme and the EC scheme cannot achieve secure transmission when caching capacity is very low as marked by the purple circle in Fig. \ref{fig1011:mos}. (a). The MOS values of the EC-VE scheme and the VE scheme are the same at the point marked by the black circle in Fig. \ref{fig1011:mos}. (a) when caching capacity increases. This is because all encoding packets of video files are cached on EC server when caching capacity is enough. For the VE scheme, all video files are assumed to be cached on the EC server to ensure secure transmission because caching strategy cannot be adjusted. The MOS superiority of the VE scheme to the EC scheme also proves the advantage of encoding strategy compared with caching strategy.

Fig. \ref{fig1011:mos}. (b) shows the average MOS versus different number of video files with $n=10$, $K=1$ and ${\Phi _k}=39000\text{Mb},\forall k \in {\cal{K}}$. As marked by the black circles in Fig. \ref{fig1011:mos}. (b), when the number of video files is 4 and 8, the average MOS values of the EC-VE scheme and the VE scheme decrease with the increasing number of  video files, while the average MOS of the EC scheme remains stable. The EC-VE scheme and the VE scheme have the same performance, which is significantly superior to that of the EC scheme. This is because the EC-VE scheme and the VE scheme achieve the highest average MOS values with the maximum encoding rates when the caching capacity is sufficient to cache all encoding packets on EC server. The EC scheme shows the constant average MOS value with all video files being encoded into the minimum encoding rates. When the number of video files is more than 8, only the EC-VE scheme can ensure secure transmission of all video files by caching less encoding packets of each video file and encoding each video file into lower encoding rate under limited caching capacity. The EC scheme and the VE scheme solely depending on caching strategy or encoding strategy cannot find feasible solutions when the number of video files exceeds 8.

\begin{figure}[!t]
	\centering
	\includegraphics[width=80mm]{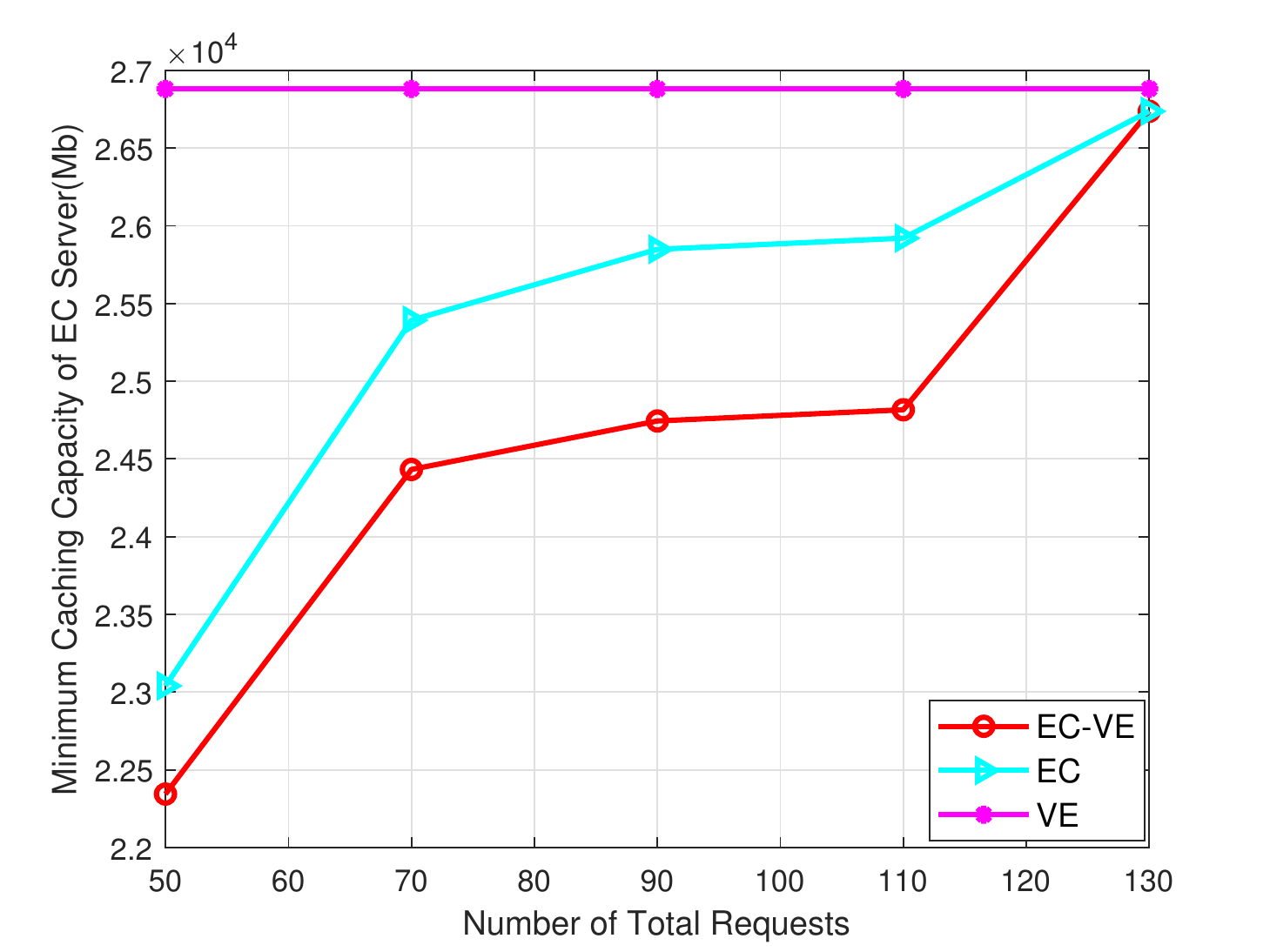}
	\caption{Minimum caching capacity vs. different number of total requests.}
	\label{fig12:mos}
\end{figure}

Fig. \ref{fig12:mos} shows the minimum caching capacity of the EC server versus different number of total requests with $n=10$, $F=8$, $K=1$. The minimum caching capacity of the EC-VE scheme and the EC scheme increases with the number of video total requests and reaches the same value at the last point. The EC-VE scheme requires much less caching capacity to ensure security than the EC scheme. Even if the minimum encoding rates are adopted by the EC scheme to obtain the minimum caching capacity to ensure security, the minimum caching capacity is still significantly more than that of the EC-VE scheme. This further demonstrates the advantage of joint optimization of caching and encoding strategy for EC-VE scheme. Meanwhile, the minimum caching capacity of the VE scheme keeps constant as the number of total requests increases because all encoding packets are cached on the EC server to ensure security. This demands at least caching capacity of ${\Phi _k}=26880\text{Mb}$ with each video file being encoded into the minimum encoding rate. 

From the analysis above, we can know that the edge caching module is responsible for caching enough packets with as small caching capacity as possible to ensure secure transmission and sparing as large caching capacity as possible for the video encoding module to obtain high video encoding quality, while the video encoding module is responsible for encoding video files into as high encoding rates as possible under secure transmission. When the edge caching module acts alone and the encoding rates of video files cannot be adjusted, the system performance is poor even if the caching capacity is sufficient. Similarly, when the video encoding module acts alone and the caching strategy cannot be adjusted, the limited caching capacity is used to ensure security and there is no enough margin to improve video encoding quality. Therefore, it is necessary to jointly optimize edge caching module and video encoding module because they interact with each other and cooperatively act on QoE improvement and security guarantee. 

\subsection{Complexity Analysis}\label{sub2}
In this subsection, the computational complexity of our proposed EC-VE algorithm, Greedy EC-VE algorithm, the ECST scheme in \cite{Gabry2016icc}, the EC-BF scheme in \cite{haoiot2020} and the EC-TM scheme in \cite{xu2019itj} is analyzed.
 
(1) EC-VE: For our proposed near-optimal EC-VE algorithm, the computational complexity is determined by the GlobalSearch algorithm and 0-1 branch and bound method as shown in Algorithm 1. For the GlobalSearch algorithm, the dimension of the optimization variable is ${\left( {K + 1} \right) \cdot n \cdot F}$. Then, the computational complexity of the GlobalSearch algorithm can be given by $O({{h_g}\left( {\left( {K + 1} \right) \cdot n \cdot F} \right)})$, where ${h_g}\left(  \cdot  \right)$ is the number of operations positively related to the dimension of the optimization variable. Finally, the computational complexity of the EC-VE algorithm is $O\left( {{h_g}\left( {\left( {K + 1} \right) \cdot n \cdot F} \right) + {N_n} \cdot {N_b} \cdot {h_g}\left( {\left( {K + 1} \right) \cdot n \cdot F} \right)} \right)$.  

(2) Greedy EC-VE: As shown in Algorithm 2, the computational complexity of our proposed Greedy EC-VE algorithm is determined by the Algorithm 1 with a fixed known ${\bf{S}}$ and the optimization algorithm of encoding parameters based on the optimized caching strategy ${{\bf{\tilde M}}_g}$. For Algorithm 1 with a fixed known ${\bf{S}}$, since the dimension of optimization variable is ${K \cdot n \cdot F}$, then the computational complexity can be given by $O({h_g}\left( {K \cdot n \cdot F} \right))$. Finally, the computational complexity of the Greedy EC-VE algorithm is $O\left( {{h_g}\left( {K \cdot n \cdot F} \right) + {N_n} \cdot {N_b} \cdot {h_g}\left( {K \cdot n \cdot F} \right)}+N_1\cdot N_2 \right)$.

(3) ECST and EC-TM: For the ECST scheme and EC-TM scheme, since they only aim to optimize caching strategy without considering video encoding, then the computational complexity is $O\left( {{h_g}\left( {K \cdot n \cdot F} \right) + {N_n} \cdot {N_b} \cdot {h_g}\left( {K \cdot n \cdot F} \right)} \right)$.

(4) EC-BF: For the EC-BF scheme, since it aims to optimize caching strategy and secure transmission rate without considering video encoding, then the computational complexity is
$O\left( {{h_g}\left( ({K \cdot n} + 1\right)\cdot F) + {N_n} \cdot {N_b} \cdot {h_g}\left( ({K \cdot n} +1)\cdot F\right)} \right)$.

Based on the analysis above, compared with the Greedy EC-VE algorithm, the increase in the dimension of the optimization variable in our proposed near-optimal EC-VE algorithm brings higher computational complexity. It has a much greater impact on the running time than the increase in the number of iterations caused by optimizing the encoding parameters alone based on the greedy method. In our simulation, the running time of the near-optimal EC-VE algorithm is approximately ${30}$ times the running time of the Greedy EC-VE algorithm. For the ECST in \cite{Gabry2016icc} and the EC-TM scheme in \cite{xu2019itj}, the computational complexity is lower than that of our Greedy EC-VE algorithm because the video encoding is not considered. However, as mentioned above, the increase in the number of iterations for optimizing the encoding parameters has a trivial effect on the running time. Therefore, the difference of computational complexity among the Greedy EC-VE algorithm, the ECST scheme and the EC-TM scheme is very small and can be considered almost the same to some extent. For the EC-BF scheme in \cite{haoiot2020}, the computational complexity is higher than that of Greedy EC-VE scheme, ECST scheme and EC-TM scheme due to the fact that it aims to optimize caching strategy and secure transmission rate simultaneously.

\section{Conclusion}
In this paper, we proposed a QoE-driven cross-layer optimization scheme for secure video transmission over the backhaul links in cloud-edge networks. First, we developed a secure transmission model based on video encoding and edge caching. By employing this model as the security constraint and considering the interaction of video encoding parameters and edge caching strategy, we formulated a joint optimization problem of video encoding parameters and edge caching strategy to improve QoE. Then, a near-optimal algorithm was designed to solve the joint optimization problem. Furthermore, we proposed a greedy algorithm with low complexity to obtain the suboptimal solution. Simulation results were presented to show that our proposed algorithms can greatly improve video encoding quality and reduce transmission latency under the condition of ensuring secure transmission of the backhaul links compared with the existing algorithms. In addition, our proposed algorithms were proven to be more robust for caching capacity and could ensure secure transmission for more videos with limited caching capacity of edge caching servers. 


%

\begin{IEEEbiography}[{\includegraphics[width=1.0in,height=1.25in,clip,keepaspectratio]{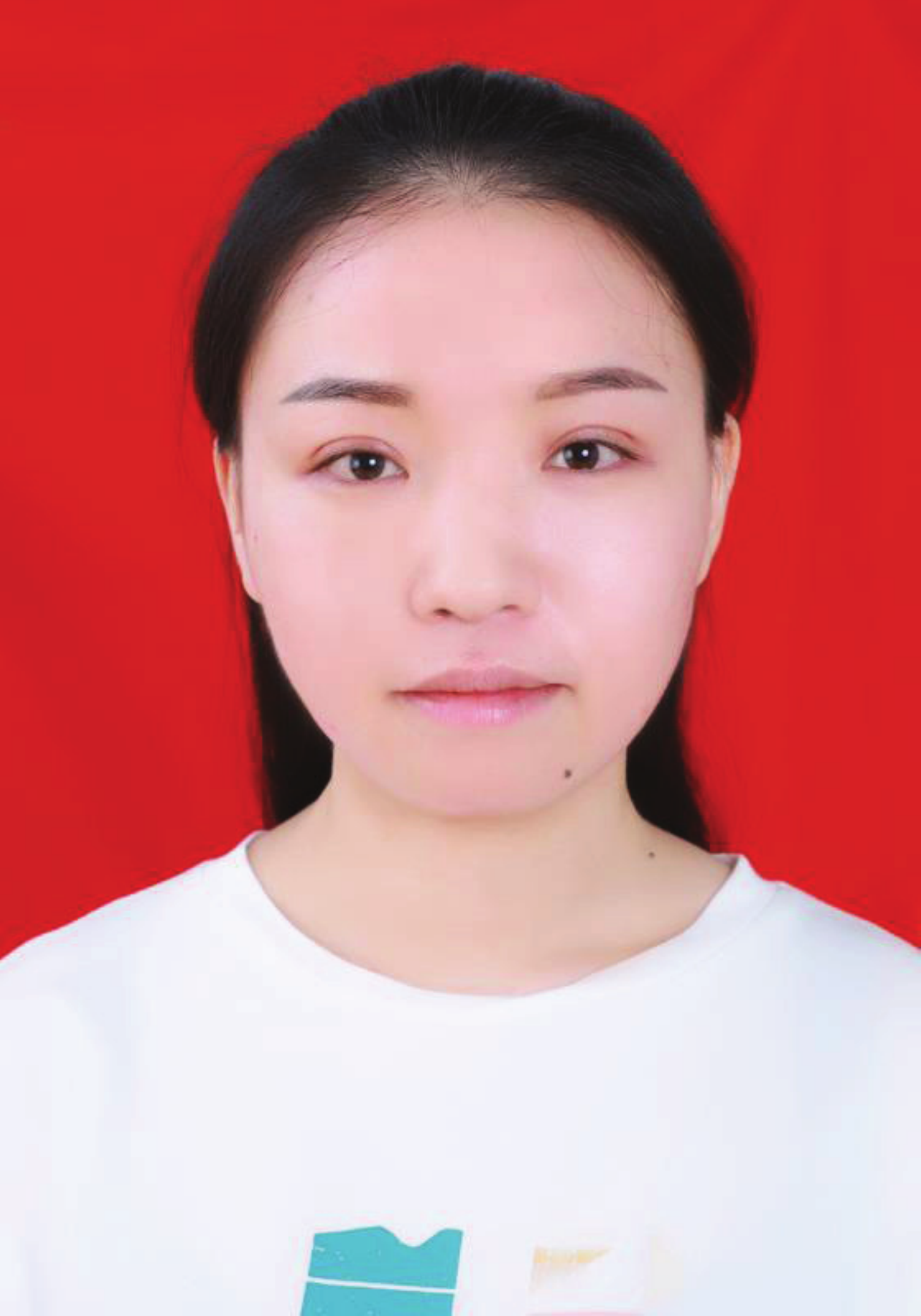}}]{Tantan Zhao}	
received the B.S. degree in communication engineering from Shanghai University, Shanghai, China, in 2014, and the M.S. degree in information and communications engineering from Xi'an Jiaotong University, Xi'an, China, in 2018. She is currently pursuing the Ph.D. degree with the School of Information and Communications Engineering, Xi'an Jiaotong University, China. Her research interests include video communication and transmission, mobile edge computing and deep reinforcement learning.
\end{IEEEbiography}
\vspace{-2.8cm}
\begin{IEEEbiography}[{\includegraphics[width=1.0in,height=1.25in,clip,keepaspectratio]{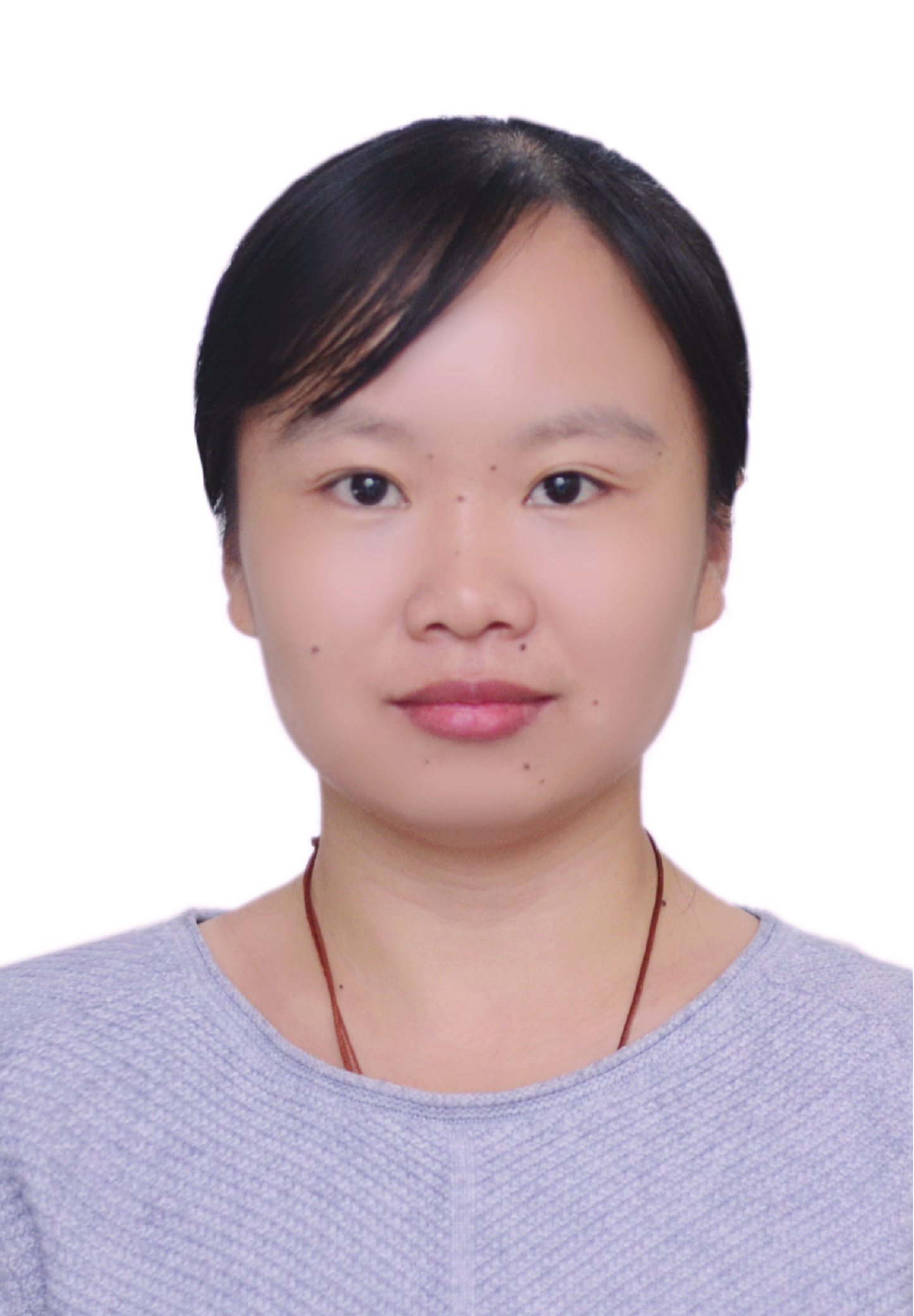}}]{Lijun He}	
received the B.S. and Ph.D. degrees from the School of Information and Communications Engineering, Xi'an Jiaotong University, Xi'an, China, in 2008 and 2016, respectively. She is currently a Associate Professor with the School of Information and Communications Engineering, Xi'an Jiaotong University. Her research interests include video communication and transmission, video analysis, processing, and compression techniques.
\end{IEEEbiography}
\vspace{-2.8cm}
\begin{IEEEbiography}[{\includegraphics[width=1.0in,height=1.25in,clip,keepaspectratio]{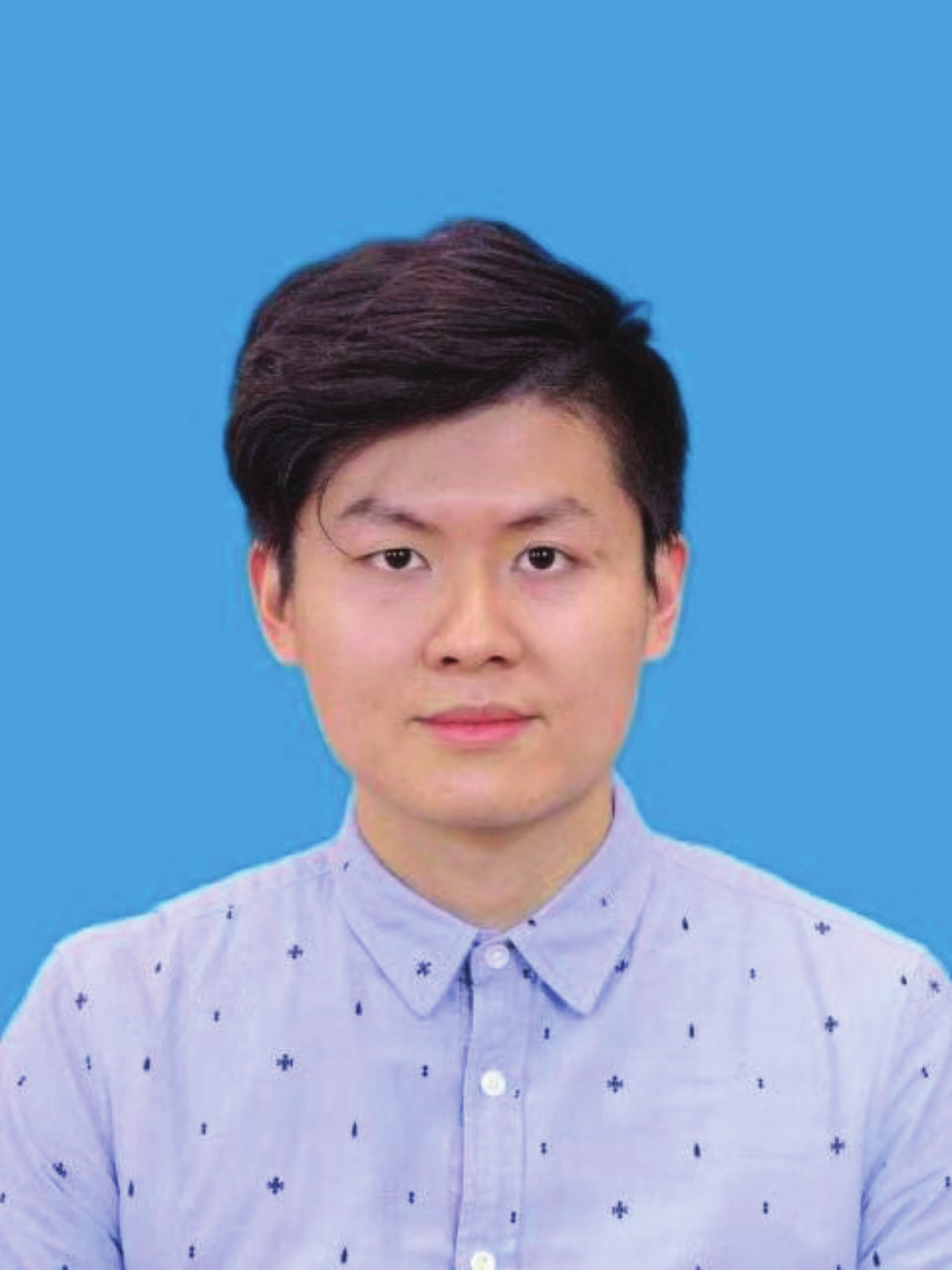}}]{Xinyu Huang}	
received the B.S. degree from the Tsien Elite Class in the Xidian University, in 2018. He is currently pursuing the M.S. degree on the Information and Communications Engineering in the Xi'an Jiaotong University. His research interests include edge computing, video transmission and deep learning.
\end{IEEEbiography}
\vspace{-2.8cm}
\begin{IEEEbiography}[{\includegraphics[width=1.0in,height=1.25in,clip,keepaspectratio]{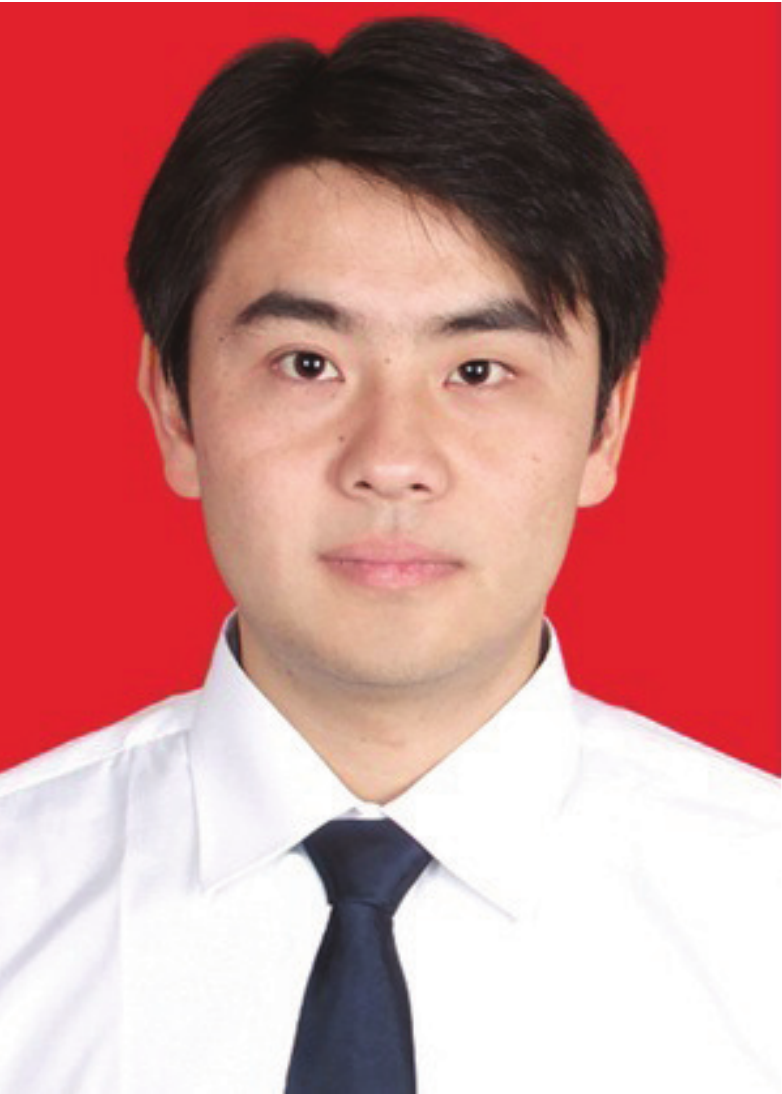}}]{Fan Li}	
(M'10) obtained his B.S. and Ph.D. degrees in information engineering from Xi'an Jiaotong University, Xi'an, China, in 2003 and 2010, respectively. From 2017 to 2018, he was a Visiting Scholar with the Department of Electrical and Computer Engineering, University of California, San Diego. He is currently a Professor with the School of Information and Communications Engineering, Xi'an Jiaotong University. He has published more than 60 technical papers. His research interests include multimedia communication, image/video coding and image/video quality assessment. He served as the Local Chair for ICST Wicon 2011, and was a member of the Organizing Committee for IET VIE 2008.
\end{IEEEbiography}

\end{document}